\documentclass[preprint,12pt, 3p]{elsarticle}
\usepackage{graphicx}
\usepackage{latexsym}
\usepackage[utf8]{inputenc}
\usepackage{amsthm}
\usepackage{mathtools}  
\usepackage{amsmath}
\usepackage{amssymb}
\usepackage{amsfonts}
\usepackage{multirow}
\usepackage{caption,subcaption}
\usepackage{url}
\usepackage{float}
\usepackage{xcolor}
\usepackage{booktabs}
\usepackage{adjustbox}
\usepackage{soul}
\usepackage{lineno}
\usepackage{caption,subcaption}
\usepackage{setspace}
\usepackage{fixmath}
\usepackage{adjustbox}
\usepackage{array} 

\usepackage[ruled,algonl,vlined]{algorithm2e}
\usepackage{lineno}

\usepackage{lineno}

\usepackage{lineno}
\newcommand{\adamya}[1]{{#1}}
\graphicspath{{Figures/}}

\begin{document}

\begin{frontmatter}
	\title{UniRecSys:  A Unified Framework for Personalized, Group, Package, and Package-to-Group Recommendations}

    \author[1]{Adamya Shyam}
    \ead{ashyam@cs.du.ac.in}
    \author[1]{Vikas Kumar\corref{cor1}}
    \ead{vikas007bca@gmail.com}
    
   \author[2]{Venkateswara Rao Kagita\corref{cor1}}
   \ead{venkat.kagita@nitw.ac.in}
    \author[3]{Arun K Pujari}
    \ead{arun.pujari@mahindrauniversity.edu.in}
    
    \address[1]{University of Delhi, Delhi, India}
    \address[2]{National Institute of Technology, Warangal, India}

    \address[3]{Mahindra University, Hyderabad, India}
     \cortext[cor1]{Corresponding author}

\begin{abstract}

Recommender systems aim to enhance the overall user experience by providing tailored recommendations for a variety of products and services. These systems help users make more informed decisions, leading to greater user engagement with the platform. However, the implementation of these systems largely depends on the context, which can vary from recommending an item or package to a user or a group. This requires careful exploration of several models during the deployment, as there is no comprehensive and unified approach that deals with recommendations at different levels. Furthermore, these individual models must be closely attuned to their generated recommendations depending on the context to prevent significant variation in their generated recommendations. In this paper, we propose a novel unified recommendation framework that addresses all four recommendation tasks, namely, personalized, group, package, and package-to-group recommendation, filling the gap in the current research landscape. The proposed framework can be integrated with most of the traditional matrix factorization-based collaborative filtering (CF) models. This research underscores the significance of including group and package information while learning latent representations of users and items for personalized recommendations. These components help in exploiting a rich latent representation of the user/item by enforcing them to align closely with their corresponding group/package representation. We consider two prominent CF techniques, namely Regularized Matrix Factorization and Maximum Margin Matrix factorization, as the baseline models and demonstrate their customization to various recommendation tasks. Experimental results on two publicly available datasets are reported, comparing them to other baseline approaches for various recommendation tasks.

\end{abstract}


\end{frontmatter}


\section{Introduction} 

Recommender Systems (RS) are a subset of information-filtering systems that helps users to make an informed decision about the product or services of their interest, with diverse applications in several domains, including e-commerce~\cite{schafer1999recommender, sarwar2002recommender, gu2020hierarchical}, social networks~\cite{rivas2020social, seo2017personalized}, e-learning~\cite{tarus2018knowledge, rahayu2022systematic}, and tourism~\cite{borras2014intelligent, esmaeili2020novel}. With the rapid explosion in digital information available on various e-commerce and social media platforms, recommender systems have been extensively adopted in many business applications to effectively retrieve valuable information from arbitrarily large amounts of data~\cite{resnick1997recommender, ricci2022recommender}. The defining characteristic of any RS algorithm is personalization, which requires exploiting user interests over items based on past interactions, such as ratings or reviews. RS can be customized to generate personalized suggestions for an individual or group of users about an item or a group of items (package), leading to four possible recommendation scenarios: personalized recommendation, group recommendation, package recommendation, and package-to-group recommendation. Personalized recommendation algorithms are designed to suggest products or services to individual users based on their past preferences, while group recommendation algorithms recommend items to groups based on the individual preferences of their members. In a package recommendation, items are recommended together as a package to individual users, and package-to-group recommendation suggests packages to groups by exploiting groups' preference towards packages.


Several approaches have been proposed in the literature to tackle diverse recommendation scenarios, considering each category as an independent optimization task. To the best of our knowledge, no approach handles various recommendation scenarios as a unified optimization problem and allows customization to any scenario depending on the requirements. In this paper, we propose a generic unified framework that can be integrated with most of the traditional matrix factorization (MF) based collaborative filtering algorithms, effectively leveraging all four possible recommendation scenarios. We consider MF-based models due to their popularity among several categories of recommendation tasks. These models have been shown to outperform traditional CF models in terms of recommendation accuracy. Usually, MF-based approaches assume that users (groups) and items (packages) can be represented in a latent space, and their interaction can be easily modeled as the inner product of their latent representation minimizing an explicit or implicit loss function. Existing MF approaches either focus on extracting users/items representation or groups/packages representation independently, which makes them inefficient in customizing for different recommendation tasks. Given the preference matrix, our proposal simultaneously learns the latent factor representations for users, items, groups, and packages that can later be used for generating preferences of a user (group) over items (packages). In a personalized setting, preference for a user over items can be generated by taking the inner product between their feature representations. We also explore various ways of generating preferences for the recommendation task involving groups and/or packages. For example, the preference for a group over items can be predicted using the latent representation of the group and the items or by performing aggregation over the individual user (item) preferences within a group (package).

\adamya{The proposed unified recommendation framework is poised to find widespread application across industries such as e-commerce, entertainment, tourism, and social media. Its implementation enables these sectors to deliver personalized and diverse recommendations to users, hence enhancing user engagement and loyalty. For example, consider a group of friends planning a vacation to a tourist destination. Each group member has specific activities, attractions, or accommodations preferences that may not align entirely with others. One may prefer adventure activities, while another may prefer cultural events. Hence, the recommendation module must have a provision to be customized for the personalized recommendation task. Certain activities can be enjoyed collectively by the entire group, such as guided tours and group excursions, and hence require customization at the group level. Similarly, a group member may only be interested in some trip activities and wish to opt for a predefined package such as adventure packages, luxury packages, or budget-friendly options. In some cases, a few of the friends from the group may consider a set of activities together depending on the group discount, exclusive group activities, or accommodations. Presently, such scenarios are prevalent in numerous online platforms. For instance, music streaming website \textit{Spotify}\footnote{https://open.spotify.com} not only provides personalized recommendations but also has a feature called \textit{Blend}, which provides recommendations for friends. Similarly, \textit{Teleparty}\footnote{https://www.teleparty.com/} is an extension that enables groups to watch movies on streaming websites such as Netflix\footnote{https://www.netflix.com}, HBO\footnote{https://www.hbo.com/}, etc. }

\noindent 
\adamya{The main contributions of this paper can be summarized as follows:}
\begin{itemize}
    \item \adamya{We propose a unified framework that can be integrated with any traditional matrix factorization-based collaborative filtering model for various recommendation tasks, namely personalized recommendation, group recommendation, package recommendation, and package-to-group recommendation.}
    
    \item \adamya{The proposed framework emphasizes the importance of incorporating group and package information while making personalized recommendations. The group and package information can be used as an anchor point to guide the learning process, focusing on exploiting user and item representation.}
    
    \item \adamya{Our proposal simultaneously learns the latent representation of users, items, groups, and packages, making it easy to customize further for specific recommendation tasks.} 
    
    \item \adamya{We show through extensive experimentation that our proposal enhances the accuracy of all four categories of recommendation tasks.}

\end{itemize} 

\adamya{The proposed approach not only streamlines the recommendation process but also covers a spectrum of stakeholders by providing personalized and group recommendations (user-centered) as well as package and package-to-group recommendations (provider-centered), treating both stakeholders fairly~\cite{Amigo2023Utility}. We show through extensive experimentation that our proposal enhances the accuracy of all four categories of recommendation tasks.}

The remainder of this paper is organized as follows. Section~\ref{Related Work} briefly discusses the four categories of recommendation tasks. In Section~\ref{MFCF}, we present an introductory discussion on matrix factorization and its corresponding formulation of optimization problems in the context of recommendation systems. We introduce
our proposed method in Section~\ref{UMFFR}. In Section~\ref{expSec}, our
methods are evaluated on two publicly available movie rating datasets. Finally, in Section~\ref{sec:conclusion}, we present the conclusion and future work. 

\section{Related Work}
\label{Related Work}
Since there is no previous work on a unified framework addressing four different categories of recommendation scenarios, namely personalized recommendation, group recommendation, package recommendation, and package-to-group recommendation, in this section, we briefly review some related work for each recommendation scenario.

\subsection{Personalised Recommendation}
Personalized recommender systems (PRS) are one of the most widely used categories
of RS that exploit the feedback provided by the user to uncover their latent interests in
items~\cite{LU20121}. There are two main approaches to PRS, namely content-based filtering (CBF) and collaborative filtering (CF). CBF relies on the similarity score between the content information used for describing the users and items to produce recommendations~\cite{park2012literature}. One major limitation of the CBF approach is that the recommendations often lack diversity and contain items similar to those a user has already interacted with~\cite{blanco2008flexible}. In contrast, the CF approach recommends items to a user based on her past preferences and the preference information of other similar users~\cite{su2009survey}. Among several approaches of CF,  the matrix factorization (MF) based approach is most widely adopted\adamya{,} primarily \adamya{due} to its effectiveness in handling the data sparsity. Here we provide a brief review of the MF-based personalized recommendation approaches, a core ingredient of the proposed framework.

The introduction of matrix factorization approaches in the Netflix Prize competition proved that matrix factorization performs superior over traditional algorithms and also can deal with the problem of data scarcity~\cite{bennet2007netflix}. Regularized matrix factorization (RMF) minimizes a cost function quantifying the discrepancy between the observed ratings and their corresponding prediction using the squared loss~\cite{wu2007collaborative}. A regularization component over the latent representation of users and items is also included in the cost function to avoid overfitting. The non-negative matrix factorization (NMF) approach of CF assimilates the non-negativity constraint enhancing the interpretability of the latent representation of the user and items. In other words, NMF considers the fact that most of the real-world data are non-negative and maintains such constraint in the factorization~\cite{lee2000algorithms}. Mnih et al.~\cite{mnih2007probabilistic} proposed a probabilistic matrix factorization (PMF) approach that assumes that the observed ratings follow a normal distribution and \adamya{proposes} to maximize the log-posterior over the movie and user features. The PMF model includes priors over the user and item feature vectors for automatic control of model complexity and to avoid overfitting and hence, generalizing better for new data. Rennie et al.~\cite{rennie2005fast} proposed a maximum margin matrix factorization (MMMF) formulation for CF over the rating matrix. MMMF envisages the rating prediction task as a multi-class classification problem and modifies the hinge loss, a popular loss function for margin maximization, for the ordinal rating matrix factorization setting.  MMMF has become a very popular research topic since its publication, and several extensions have been proposed~\cite{kumar2017collaborative, weimer2007cofi, decoste2006collaborative, xu2012nonparametric,kumar2017proximal, xu2013fast}. In~\cite{kumar2017proximal}, a proximity criterion is proposed as an alternative to margin maximization. Xue et al.~\cite{xue2017deep} proposed a novel MF approach with a deep neural network architecture that uses a binary cross entropy to model the non-linear interactions. In~\cite{yi2019deep}, deep matrix factorization (DMF) is proposed to enhance the representation of users and items based on implicit feedback and various \adamya{kinds} of side information.

\subsection{Group Recommendation}
Group recommendation systems (GRS) aim to construct a model for group preference prediction by exploiting the past preferences of the group members over items. Several approaches have been proposed in the literature, which can be broadly classified into two categories viz. profile aggregation and recommendation aggregation. In the first approach, the group members' individual profiles are aggregated to construct a group profile and later the items are recommended based on the group profile~\cite{shi2015latent, kagita2015virtual,kagita2013precedence,kagita2013group}. In the recommendation aggregation approach, the personalized recommendations of the group members are aggregated in order to provide group recommendations~\cite{kim2015stochastic}. The aggregation for these approaches can be achieved using various techniques such as average~\cite{mccarthy2002pocket}, weighted average~\cite{felfernig2018algorithms, yalcin2021novel}, least misery~\cite{ortega2016recommending, nawi2020evaluation, o2001polylens}, and most pleasure~\cite{senot2010analysis, dara2020survey}.  
 
In \cite{cao2018attentive} a neural collaborative filtering (NCF) based approach was proposed to adapt the group representation using attention mechanism and learn the group-item interactions. In a similar work~\cite{yin2019social}, a bipartite graph embedding technique is integrated with an attention mechanism to learn user embedding as well as user's social influences in a unified way and predict group-item interactions. Nozari et al.~\cite{nozari2020grs} proposed to model the influence of members and leaders of groups to enhance the group recommendation quality. \adamya{Acharya et al.~\cite{Acharya2023Utility} proposed to use spatial binning to group venues in a user's proximity based on their past history to achieve point-of-interest (POI) based recommendation.} In a latent group model (LGM)~\cite{shi2015latent} approach, the authors detected the groups using an automated fashion by applying the k-means algorithm on the user latent factor matrix obtained using matrix factorization. The latent factors of group members were further aggregated to construct the group profile and provide group recommendations using matrix factorization. Extending the matrix factorization model, Jeong et al.~\cite{jeong2019hggc} took into account the idea of group cohesion, and proposed a hybrid model incorporating content information and rating data simultaneously. In several other studies, matrix factorization-based extensions for GRS \adamya{have} been proposed \cite{bobadilla2017recommender, liu2018collaborative, wang2021group, davtalab2021poi}. In recent studies, the use of deep neural networks has also been done to model group preferences and recommendation purposes \cite{huang2020efficient, wang2019novel}.

\subsection{Package Recommendation}
The \adamya{task} of a package recommendation system, also known as a bundle recommender\adamya{,} is to provide a list of items (package) to a user as the recommendation. 
These types of systems prove to be very significant in domains like travel~\cite{xie2012composite, ge2014cost}, music~\cite{cao2017embedding}, and clothing~\cite{wibowo2018incorporating} where a bundle of items will be more preferred over a single item. For example, in travel recommendations, a list of tourist spots can be bundled together to provide an ideal route for the user. To develop a package recommender, generally, the pipeline consists of bundle generation, model building, and then the prediction task. Mostly, the bundle generation task has been viewed as either a knapsack problem~\cite{yang2012tourist, li2015exploring, yu2015personalized, van2019package} or a clustering problem~\cite{mengash2014gcar, qi2018recommending, ortiz2019clustering}, few of the studies have also used predefined bundles~\cite{tan2014object, sharma2019learning, alsayasneh2017personalized}. 

Viewing the problem as a structured prediction task, Bai et al.~\cite{bai2019personalized} proposed a bundle generation network (BGN) that employs an encoder-decoder framework with a feature-aware softmax and generates a bundle using masked beam search. In~\cite{zhu2021neural}, a Neural Attentive Travel Package Recommendation (NATR) is proposed. The framework leverages both long-term and short-term user behaviors to provide recommendations. NATR utilizes a neural network with attention mechanisms to capture the user's historical preferences, recent interests, and the influence of social connections. While studies have also benefited from the deep learning methods~\cite{wang2019neural, he2017neural, chen2019matching, cao2017embedding}, package recommender systems commonly use collaborative filtering methods like matrix factorization. In~\cite{pathak2017generating}, bundle rankings were modeled by comprising the parameters learned through the item Bayesian Personalized Ranking (BPR) to bundle BPR for existing bundles. Further, the learned parameters were also used to create personalized bundles. Wibowo et al.~\cite{wibowo2017matrix} created item constraints in two different ways, handcrafted and automated; and extended the matrix factorization technique to provide package recommendations by incorporating these constraints in the learning of user and item embeddings. 

\subsection{Package-to-Group Recommendation}
The diverse application of recommender systems in different scenarios has led to the development of \adamya{Package-to-Group (P2G)} recommendation models where a list of items has to be recommended to a group of users.  For example, within the tourism sector, a travel agency might have to suggest vacation packages tailored for specific group sizes or demographics. 
Although recommendation techniques have seen significant progress, only a handful of studies have focused on the 
P2G recommendation task. Some of these studies have introduced probabilistic models to address this task. 
\adamya{Given a group of users with individual preferences of each group member, these approaches estimate the probability score to measure the likelihood of group G selecting the package P.} 
The estimated score is dependent on the individual preference of each group member for each item of the package. The first approach estimates the probability of an item being selected by the group. Further, it combines the individual scores to estimate the package score, and the second works reversely by first estimating the probability score of a package being selected by each group member and further aggregates them to derive package-to-group score \cite{benouaret2018package, qi2016recommending, qi2018recommending}. Apart from this, most of the work done in this area focuses on the fairness of the packages and groups used for the task \cite{serbos2017fairness, sato2022enumerating}. 

\adamya{Table~\ref{tab:related_works} provides a comprehensive review of papers covered under each category of recommendation task.
Most of the traditional recommendation algorithms proposed so far model the optimization problem for each category of recommendation task separately and hence fail to capture intrinsic relationships among different recommendation categories.} 
In this paper, we have proposed a unified recommender system based on the matrix factorization technique that models the users' as well as groups' preferences for items and packages. 

\begin{table}[t]
    \renewcommand*{\arraystretch}{1.2}
    \centering
    \caption{\adamya{Summary of related research under each category of recommendation task.}}
    \begin{tabular}{lcl}
    \toprule
        Recommendation Task &  & Methods \\ \hline
        Personalized & : & \cite{bennet2007netflix},~\cite{wu2007collaborative},~\cite{lee2000algorithms},~\cite{mnih2007probabilistic},~\cite{rennie2005fast},~\cite{kumar2017collaborative},~\cite{weimer2007cofi},~\cite{decoste2006collaborative},~\cite{xu2012nonparametric},~\cite{kumar2017proximal},~\cite{xu2013fast},~\cite{xue2017deep},
        \\
        &   & \cite{yi2019deep}
        \\
        Group & : & \cite{shi2015latent},~\cite{kagita2015virtual},~\cite{kagita2013precedence},~\cite{kagita2013group},~\cite{kim2015stochastic},~\cite{mccarthy2002pocket},~\cite{felfernig2018algorithms},~\cite{yalcin2021novel},~\cite{ortega2016recommending},~\cite{o2001polylens},~\cite{senot2010analysis},~\cite{cao2018attentive},        
        \\ &   & \cite{yin2019social},~\cite{nozari2020grs},~\cite{jeong2019hggc},~\cite{liu2018collaborative},~\cite{wang2021group},~\cite{Acharya2023Utility},~\cite{huang2020efficient}
        \\
        Package & : & \cite{xie2012composite},~\cite{ge2014cost},~\cite{cao2017embedding},~\cite{wibowo2018incorporating},~\cite{yang2012tourist},~\cite{li2015exploring},~\cite{yu2015personalized},~\cite{ortiz2019clustering},~\cite{tan2014object},~\cite{sharma2019learning},~\cite{bai2019personalized},~\cite{zhu2021neural},
        \\
        &   & \cite{chen2019matching},~\cite{pathak2017generating},~\cite{wibowo2017matrix}
        \\
        Package-to-Group & : & \cite{qi2018recommending},~\cite{benouaret2018package},~\cite{qi2016recommending},~\cite{serbos2017fairness},~\cite{sato2022enumerating}
        \\
    \bottomrule
    \end{tabular}
    \label{tab:related_works}
\end{table}

\section{Matrix Factorization}
\label{MFCF}
Matrix factorization (MF) is one of the most widely used techniques for collaborative filtering (CF) that produces highly efficient and accurate automated recommendations by leveraging the underlying patterns in the user-item interaction data. The underlying principle of the MF-based CF technique is to decompose a partially observed user-item interaction matrix into latent factor matrices representing users and items.  Formally, let $\mathcal{U} = \{u_1, u_2, \dots, u_n$\} be the set of $n$ users and  $\mathcal{V} = \{v_1, v_2, \dots, v_m$\} be the set of $m$ items and $Y\in \mathbb{R}^{n \times m}$ is the partially observed user-item rating matrix. Consider a set $\Omega$, which holds the index $ij$ of each observed entry in $Y = [y_{ij}]^{n \times m}$, i.e., the entry  $y_{ij} \in \{1,2,...,L\}$, if $ij \in \Omega$, otherwise $y_{ij} = 0$. Usually, the observed entries are represented by distinct ordinal numbers representing the levels from strongly dislike to strongly like.

    \begin{figure*}[th]
        \centering
        \adjustbox{max width=\textwidth}{
        \includegraphics[scale = 1]{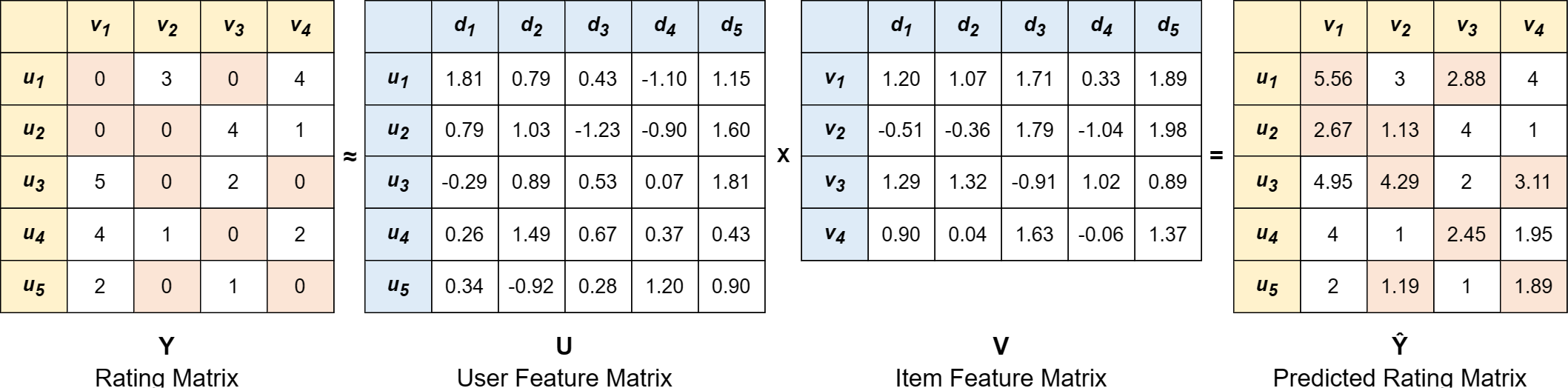}
        }
        \caption{An example of Matrix Factorization.}
        \label{fig:Matrix Factorization}
    \end{figure*}

\noindent The aim of MF-based CF model is to approximate the observed rating matrix $Y$ as the product of two latent factor matrices $U \in \mathbb{R}^{n \times d} $ and $V \in \mathbb{R}^{m \times d}$ such that $Y \approx UV^T$. Here, $d$ is a hyperparameter that denotes the size of latent space. The matrix $U$ is often referred to as the user-feature matrix, where the entries $u_{ik}$ represent the strength or influence of the $i$th user in the $k$th feature. $U_i$, the $i$th row of matrix $U$ is the latent dimension representation of $i$th user in $d$-dimensional space. Similarly, matrix $V$ is known as the item-feature matrix, and the $j$th row of $V$ denoted as $V_j$ represents the latent representation of $j$th item in the $d$-dimensional space. Typically, in most of the real-world interaction matrices, a small number of entries are observed, i.e., $\Omega \ll n\times m$, which makes the factorization nontrivial. 

\noindent Once we obtain the factor matrices representation \adamya{of} users and items, then a rating score for $i$th user over $j$th item is modeled as $\hat{y}_{ij} = \sum_{k=1}^{d}{u_{ik}v_{jk}}$ = $U_iV_j^T$. The decomposition of matrix $Y$ into latent factor $U$ and $V$ is generally achieved by solving an optimization problem consisting of a measure to quantify the loss and any additional constraints over the factor matrices.   A common formulation for this problem is a regularized loss problem written as follows.

\begin{equation}
    J(Y,U,V) = \sum_{ij \in \Omega} \Phi(y_{ij}, U_i, V_j) + \lambda \Psi(U,V),
\end{equation}

\noindent where $\lambda > 0$ is a hyperparameter for balancing the recoverability and regularization, $\Phi(\cdot)$ is a loss function that measures how well $U_iV_j^T$ approximates $y_{ij}$, $\Psi(\cdot)$ is a regularization function that promotes various desired properties in $U$ and $V$ such as sparsity and group-sparsity. There are several choices for the loss and regularization functions, yielding various new matrix factorization models. We have identified two baseline MF models, namely regularized matrix factorization (RMF)~\cite{wu2007collaborative, koren2009matrix} and maximum margin matrix factorization (MMMF)~\cite{srebro2004maximum}, to show the efficacy of the proposed approach. The motivation for selecting RMF and MMMF as baselines is based on the observation that these methods are specifically designed to handle different types of input data. RMF is a suitable choice when the user-item interaction matrix contains real-valued data, whereas MMMF is tailored made for the factorization of ordinal rating matrices.

The RMF algorithm minimizes the element-wise squared aggregated error for the observed value in order to determine a pair of factors. A regularization constraint is included to ensure that the factors do not become arbitrarily large.

\begin{equation}\label{eq2}
    \min_{U,V}J(U,V) =\sum_{\{i,j\} \in \Omega} (y_{ij} - U_{i} V_{j}^T)^2 + \frac{\lambda}{2}(\|U\|_{F}^2+\|V\|_{F}^2),
\end{equation}

\noindent where $\|A\|_{F}^2 = \sum_{ij} a_{ij}^2$. In contrast to RMF, MMMF replaces the squared loss function with a smooth version of the hinge loss, a popular loss function for margin maximization. In addition to this, it also learns $L-1$ thresholds $\{\theta_{i,1}, \theta_{i,2}, \dots,\theta_{i,L-1}\}$ for each $i$th user to relate the real-valued  $U_iV_j^T$ to the discrete $y_{ij}$, where $L$ is the maximum rating level.  

\begin{equation}\label{mmmf}
    \min_{U,V,\theta}J(U,V,\theta) = \sum_{(i,j) \in \Omega}\sum_{r = 1}^{L-1}h(T_{ij}^r(\theta_{i,r}-U_{i} V_{j}^T)) + \frac{\lambda}{2} (\|U\|_{F}^2+\|V\|_{F}^2)
\end{equation}
where	     
\[
    T_{ij}^r= 
\begin{dcases}
    +1, & \text{if }r \ge y_{ij}\\
    -1, & \text{if }r < y_{ij}
\end{dcases}
\] 

\noindent \adamya{and} $h(\cdot)$ is the smooth hinge-loss function defined as,
\[
    h(z)= 
\begin{dcases}
    0                  ,& \text{if }z \ge 1\\
    \frac{1}{2}(1-z)^2 ,& \text{if }0 < z < 1\\
    \frac{1}{2}-z      ,& \text{otherwise}
\end{dcases}
\]

\section{UMFFR: Unified Matrix Factorization based Framework for Recommender}
\label{UMFFR}
In this section, we discuss our proposed unified model that can be adopted for any of the four categories of recommendation tasks: personalized recommendation, group recommendation, package recommendation, and package-to-group recommendation. \adamya{ The proposed framework aims to jointly exploit the latent representation of the groups and packages in addition to the user's and item's latent factors.
Furthermore, for the recommendation over groups and packages, it is assumed that the users' and items' groups are available in addition to the individual preferences over items. To the best of our knowledge, there is no publicly available dataset with all possible types of preferences, i.e., individual and group preferences over items and packages. Therefore, taking a cue from recent research where the clustering approach is utilized to simulate group preferences from individual preferences~\cite{nozari2020novel, BORATTO2016165, ISMAILOGLU2022113663}, we create groups and packages with compatible users and items.} 

\subsection{\adamya{Group and Package Formation}}
\label{groupCreation}
Given the rating matrix $Y$, we have constructed the user and item groups using \textit{spectral clustering}~\cite{zelnik2004self} approach. To compute the user group, a graph $G=(\mathbb{V}, \mathbb{E})$ is first formed, where $\mathbb{V}$ represents the set of users and $\mathbb{E}$ is an edge set containing the edges between every pair of users. \adamya{Each edge $e_{ij}$ is weighted based on the similarity between users $u_i$ and $u_j$, which is calculated as the inverse of the Euclidean distance over the observed ratings.} 
Further, we have normalized the similarity value by the proportion of commonly rated items to mitigate the impact of varying user preferences. The normalization ensures that the absolute rating values do not overly influence the similarity between users with a higher proportion of common ratings. Let $O_i$ and $O_j$ hold the \adamya{indices} of items rated by the users $u_i$ and $u_j$, respectively, and $OI_{ij}$ contain the \adamya{indices} of commonly rated items. Then, the weight $w_{ij}$ for an edge $e_{ij}$ between users $u_i$ and $u_j$ is calculated as

\begin{equation}
    w_{ij} = \frac{|O_i \cap O_j | }{|O_i \cup O_j |} \times \frac{1}{1+ \sqrt{\sum_{p \in OI_{ij}} { (y_{ip} - y_{jp})^2 }  } }.
\end{equation}

\begin{figure}[ht!]
    \centering
    \includegraphics[width = 3.3in, height = 2.2in]{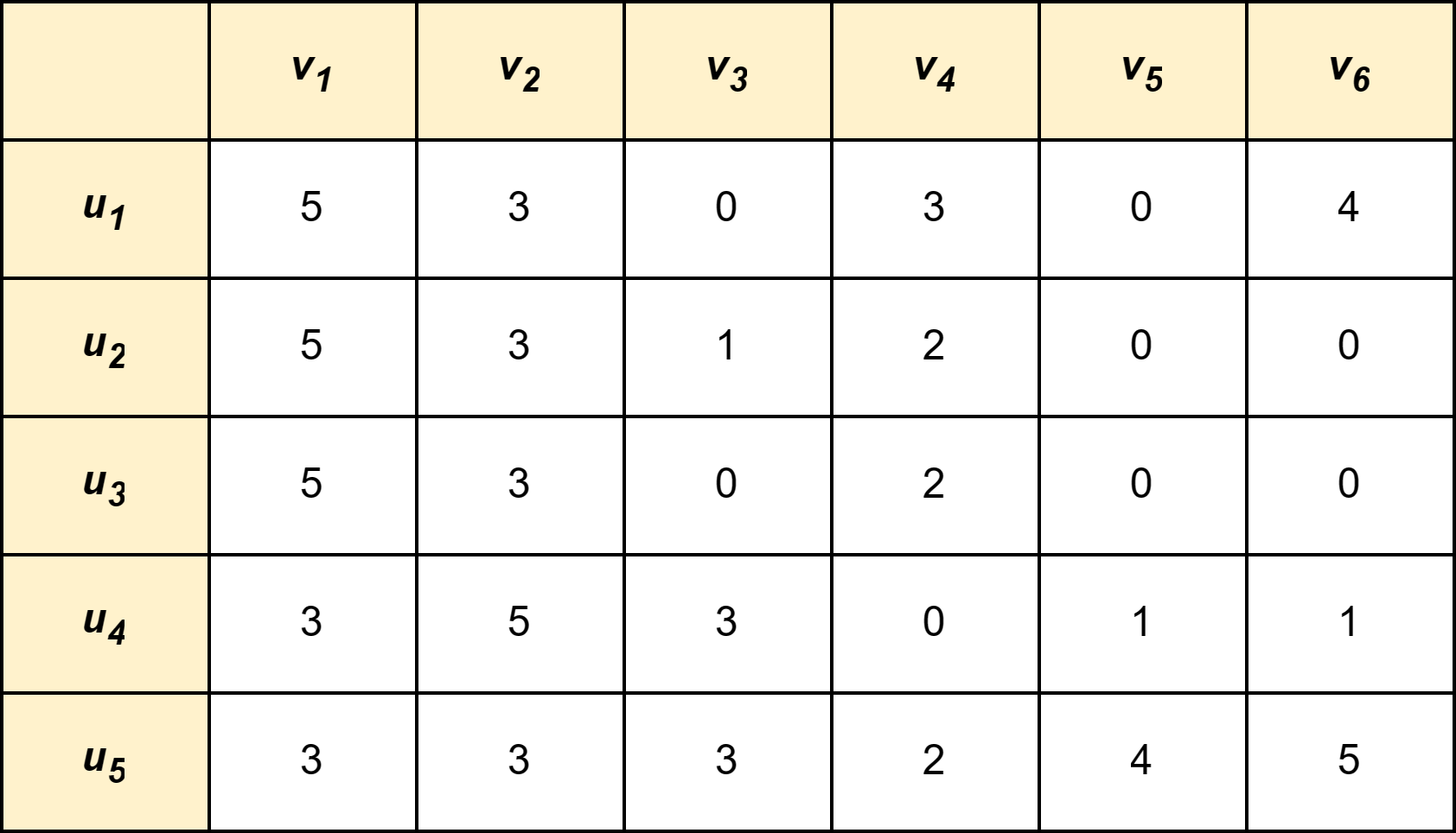}
    \caption{An example of the user-item rating matrix.}
    \label{fig:ratingMatrix}
\end{figure}

\noindent The inverse transformation ensures that the similarity value falls within the range of $[0, 1]$. A similarity value of $1$ indicates maximum similarity, while a value of $0$ indicates no similarity.  We clarify the underlying idea with the help of an example. Consider a user-item rating matrix with five users and six items (Figure~\ref{fig:ratingMatrix}). Ratings are observed on a scale of $1$-$5$, and 0's indicate the unobserved entries. The similarity between users $u_1$ and $u_2$ is calculated as

\begin{align*}
    w_{12} &=     \frac{|O_1 \cap O_2 | }{|O_1 \cup \adamya{O_2} |} \times \frac{1}{1+ \sqrt{\sum_{p \in OI_{12}} { (y_{1p} - y_{2p})^2 }  } } = \frac{3}{5} \times \frac{1}{1+ \sqrt{(5-5)^2 + (3-3)^2 + (3-2)^2} }. \\
            &= 0.6 \times 0.5 = 0.30. 
\end{align*}

\noindent Furthermore, the similarity between $u_1$ and $u_2$ is less than that between $u_1$ and $u_3$, which is $0.375$, as the measure also considers the proportion of commonly rated items as the credibility of similarity. \adamya{Subsequently, we computed the affinity matrix $A$ to preserve the graph property. Finally, the users are grouped into $K_1$ clusters by performing $k$-means with the $K_1$ largest eigenvectors of the Laplacian matrix, derived from the affinity matrix. } 
Similarly, we have formed the item groups. We do not claim that spectral clustering is the best option to obtain user and item grouping.     

\subsection{\adamya{Unified Learning of Latent Factors}}
Given the rating matrix $Y$, a set of users $\mathcal{U}$, and a set of items $\mathcal{V}$, let $(\mathcal{U}^1, \mathcal{U}^2,\dots \mathcal{U}^{K_1})$ and $(\mathcal{V}^1, \mathcal{V}^2,\dots \mathcal{V}^{K_2})$ denotes the non-overlapping grouping of user and item into $K_1$ and $K_2$ clusters, respectively, such that $\sum_{k=1}^{K_1}{|\mathcal{U}^k|} = n$ and $\sum_{k=1}^{K_2}{|\mathcal{V}^k|} = m$. The proposed generic unified framework that simultaneously learns the latent representation of users, items, user groups, and packages can be written as follows.

\begin{equation}
    J(U,V, U^G, V^G ) = \sum_{ij \in \Omega} \Phi(y_{ij}, U_i, V_j) + \lambda \Psi(U,V) + \gamma 	\Upsilon(U^G,V^G),
\end{equation}

\noindent where $\Phi(\cdot)$, $\Psi(\cdot)$ and $\lambda$ are same as defined previously. The hyperparameter $\gamma > 0$ is used to control the influence of the group information, and $\Upsilon(\cdot)$ function is induced to learn the latent factor representation of user and item groups, i.e., $U^G \in \mathbb{R}^{K_1 \times d}$ and $V^G \in \mathbb{R}^{K_2 \times d}$, respectively. $U^G_k$, the $k$th row of the matrix $U^G$ represents the latent factor representation of the $k$th user group. Similarly, the latent representation of $q$th item group ($q$th package) is represented by $V^G_q$. To learn the latent representation of user and item groups, we hypothesize that there must be a minimal deviation between the latent representation of any user and her corresponding group. Similarly, an item's latent representation must be similar to its corresponding package representation. \adamya{The incorporation of additional components ensures that the latent representation of users (items) closely aligns with their corresponding group (package) representation.}
There are several ways to impose the constraint quantifying the deviation between the user (item) and its corresponding group (package). \adamya{For our experimental setup, we minimized the sum of the squared distance between the latent representation of the user (item) and its corresponding group (package).} The modified objective function of RMF is rewritten as follows.

\adamya{\begin{align}\label{modifiedRMF}
     \min_{U, V, U^G, V^G} J(U,V, U^G, V^G ) = & \sum_{\{i,j\} \in \Omega} (y_{ij} - U_{i} V_{j}^T)^2  + \frac{\lambda}{2}(\|U\|_{F}^2+\|V\|_{F}^2) \notag \\ & + \frac{\gamma}{2} \biggl(\sum_{k=1}^{K_1}\sum_{i|i \in \mathcal{U}^k}\|U^{G}_k-U_i\|_{F}^2  + \sum_{q=1}^{K_2}\sum_{j|j \in \mathcal{V}^q}\|V^{G}_q-V_j\|_{F}^2 \biggl)        
\end{align}}

\noindent A number of optimization techniques can be employed to optimize the objective function presented in equation~\ref{modifiedRMF}. Gradient Descent method and its variants have proven to be effective, which start with random $U$, $V$, $U^G$ and $V^G$ and iteratively update $U$, $V$, $U^G$ and $V^G$ using the equations \ref{U_update}, \ref{V_update}, \ref{UG_update}, and \ref{VG_update} respectively.

\begin{align}
 U_{ip}^{t+1} &= U_{ip}^{t} - \alpha \frac{\partial J}{\partial U_{ip}^t} \label{U_update}\\
 V_{jp}^{t+1} &= V_{jp}^{t} - \alpha \frac{\partial J}{\partial V_{jp}^t} \label{V_update}\\
 U^{G^{t+1}}_{kp} &= U^{G^{t}}_{kp} - \alpha \frac{\partial J}{\partial U^{G^t}_{kp}} \label{UG_update}\\
 V^{G^{t+1}}_{qp} &= V^{G^{t}}_{qp} - \alpha \frac{\partial J}{\partial V^{G^t}_{qp}} \label{VG_update}
\end{align}

\noindent Here, the parameter $\alpha$ represents the step length, denoting the size of each step taken during the optimization process. The suffixes $t$ and $(t + 1)$ denote the current and updated values, respectively. The gradients of the variables to be optimized can be computed as follows. 

\begin{align}
\frac{\partial J}{\partial U_{ip}} &= \lambda U_{ip} - 2\sum_{\{i,j\} \in \Omega} (y_{ij} - U_{i} V_{j}^T)(V_{jp}) + \gamma \sum_{k = 1}^{K_1} \sum_{i|i \in \mathcal{U}^k}(U^{G}_k-U_i)  \\
\frac{\partial J}{\partial V_{jp}} &= \lambda V_{jp} - 2\sum_{\{i,j\} \in \Omega} (y_{ij} - U_{i} V_{j}^T)^T(U_{ip}) + \gamma \sum_{q = 1}^{K_2} \sum_{j|j \in \mathcal{V}^q}(V^{G}_q-V_j) \\
\frac{\partial J}{\partial U_{kp}^{G}} &= \gamma \sum_{k = 1}^{K_1} \sum_{i|i \in \mathcal{U}^k}(U^{G}_k-U_i)  \\
\frac{\partial J}{\partial V_{qp}^{G}} &= \gamma \sum_{q = 1}^{K_2} \sum_{j|j \in \mathcal{V}^q}(V^{G}_q-V_j)
\end{align}

\noindent Similarly, we have modified the formulation of MMMF (equation~\ref{mmmf}) by incorporating the term minimizing the deviation between the user (item) and its group (package) latent representations. As mentioned previously, MMMF also learns $R-1$ threshold for the individual user to map a real-valued prediction to a discrete rating level. Hence, we have also incorporated a term to minimize the differences between individual user thresholds and corresponding group thresholds. The unified formulation of MMMF is as follows.

\adamya{
\begin{align}\label{modifiedMMMF}
     \min_{U,V,\theta, U^G, V^G, \theta^G}J(U, V, \theta, U^G, V^G, \theta^G) =& \sum_{\{i,j\} \in \Omega}\sum_{r = 1}^{L-1}h(T_{ij}^r(\theta_{i,r}-U_{i} V_{j}^T)) + \frac{\lambda}{2} (\|U\|_{F}^2+\|V\|_{F}^2) \notag \\ & + \frac{\gamma}{2} \biggl(\sum_{k=1}^{K_1}\sum_{i|i \in \mathcal{U}^k}\|U^{G}_k-U_i\|_{F}^2  + \sum_{q=1}^{K_2}\sum_{j|j \in \mathcal{V}^q}\|V^{G}_q-V_j\|_{F}^2 \notag \\ & + \sum_{k=1}^{K_1}\sum_{i|i \in \mathcal{U}^k}\sum_{r=1}^{L-1}\|\theta^{G}_{k,r}-\theta_{i,r}\|_{F}^2 \biggl)
\end{align}}

\noindent Here $\Omega$, $\lambda$ and  $\gamma$ are same as defined previously and $\{\theta_{i,1}^G, \theta_{i,2}^G, \dots,\theta_{i,L-1}^G\}$ refers to the thresholds for the $i$th group. We follow the Gradient-based optimization approach to learn $U^G$, $V^G$, and  $\theta^G$. 

\subsection{Preference Prediction}
\label{aggStrategies}
Once $U$, $V$, $U^G$, and $V^G$ are computed, the prediction for different recommendation scenarios can be directly obtained using these latent factor matrices. Table~\ref{tab:group_ag} list preference computation for $i$th user (group) over $j$th item(package) using the latent representation. 

\begin{table}[ht!]
    \renewcommand*{\arraystretch}{1.2}
    \centering
    \caption{Preference prediction using latent factors}
    \begin{tabular}{lcl}
    \toprule
        Recommendation Task &  & Preference Prediction \\ \hline
        Personalized & : & $U_i \times V_j^T$ \\
        Group & : & $U^G_i \times V_j$ \\
        Package & : & $U_i \times V^{G^T}_{j}$ \\
        Package-to-Group & : & $U^G_{i} \times V^{G^T}_{j}$ \\
    \bottomrule
    \end{tabular}
    \label{tab:group_ag}
\end{table}

Felfernig et al.~\cite{felfernig2018evaluating} suggested a mean-based aggregation approach, where the rating for an item corresponding to a group in the test set is computed as a mean of the ratings given by users within that group. We have also performed a similar experiment to compute preferences in the situation involving group and/or package.

\begin{figure}[ht!]
    \centering
    \adjustbox{max width=\textwidth}{\includegraphics{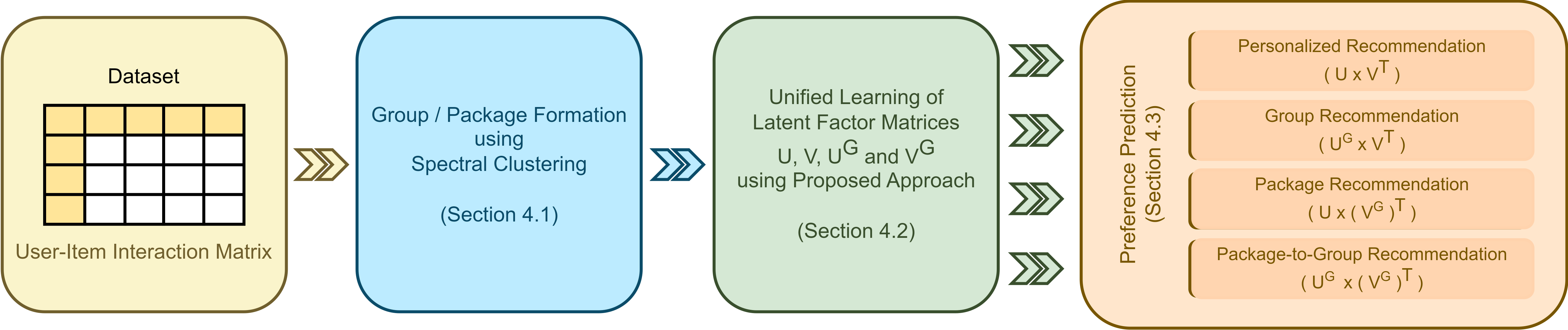}
   }
    \caption{Outline of the proposed approach.}
    \label{fig:UniRecSys}
\end{figure}

\adamya{A concise representation of our proposed framework is presented in Figure \ref{fig:UniRecSys}. Given a user-item interaction matrix $Y$, we first construct the user-user and item-item affinity matrices using the approach discussed in section 4.1. We then employ spectral clustering over the affinity matrix corresponding to users and items to create compatible groups and packages, respectively.  Subsequently, using the individual preferences over items, group, and package information, we learn latent factor matrices $(U)$, $(V)$, $(U^G)$ and $(V^G)$ containing the representation of users, items, groups, and packages, respectively.  Finally, as discussed in section~\ref{aggStrategies}, we perform the preference prediction using the estimated latent factors.}

\section{Empirical Analysis}
\label{expSec}
\adamya{This section describes the results of our experimental analysis on two publicly available real-world datasets (section~\ref{dataset}).  We examine the comparative performance of our proposed approach in terms of three offline evaluation metrics (section~\ref{sec:evalM}) against representative algorithms from each category of recommendation task (section~\ref{sec:baseline}). For a fair comparison, we also outline the detailed procedure followed to tune the hyperparameters of both the proposed and baseline algorithms (section~\ref{sec:parameterTuning}).  Subsequently, we provide the results of our extensive experimental analysis conducted across four different recommendation scenarios (section~\ref{sec:resultsandDiscussion})}.

\subsection{Datasets Preparation}
\label{dataset}
\adamya{To the best of our knowledge, there is currently no publicly available dataset encompassing all possible preference combinations, i.e., individual and group preferences
over items and packages. Hence, we utilized individual user preferences over items and then preprocessed them to simulate compatible groups and packages}. We consider two well-known benchmark datasets: MovieLens 100K\footnote{https://grouplens.org/datasets/movielens/100k/} and MovieLens 1M\footnote{https://grouplens.org/datasets/movielens/1m/}. These datasets are utilized to simulate various recommendation scenarios.  The MovieLens 100K dataset consists of $1,00,000$ ratings given by $943$ users over $1682$ movies, and MovieLens 1M dataset contains $10,00,209$ anonymous ratings given by 6,040 users for 3,706 movies. The datasets are partitioned into train and test sets for the model training and evaluation. The reported results for each recommendation scenario represent the average of three runs. We used the \textit{70:30 }train-test split ratio over the observed ratings for the personalized recommendation. In \adamya{the} group recommendation scenario, the test set contains an item along with its corresponding rating provided by different users within the group. Following this, we have included an item to the test set if it is rated by at least a user-specified percentage of the users within the group. In our experimental analysis, we consider an item as a candidate item for the test set for a particular group if it is rated by at least $20\%$ of users within that group. Further, the ratings corresponding to an item moved to the test set are set to zero in the training set for each user within the group. In addition to this, we have also set a threshold that a maximum of 5 items can be included in the test set for each group.

\begin{figure}[hbt!]
    \centering
    \adjustbox{max width=0.67\textwidth}{
    \includegraphics[scale=1]{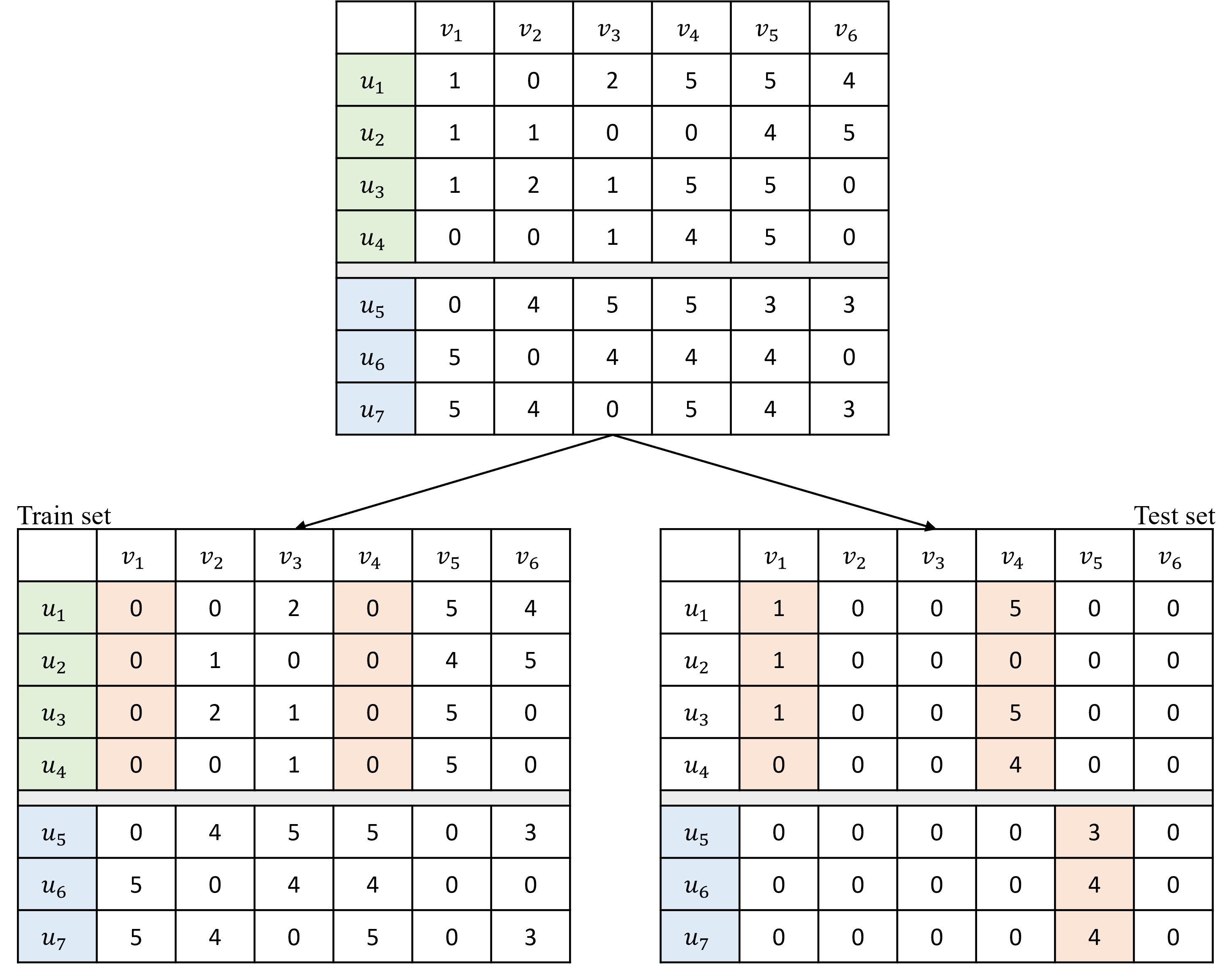}
    }
    \caption{An example of train-test split for group recommendation scenario.}
    \label{fig:testTrainSplit}
\end{figure}

\noindent We clarify the underlying train test split with the help of an example.  Consider a user-item rating matrix given in Figure~\ref{fig:testTrainSplit} with $7$ users and $6$ items. Let us assume that $\mathcal{U}^1 = \{u_1, u_2, u_3, u_4\}$ and $\mathcal{U}^2 = \{u_5, u_6, u_7\}$ are the two user groups identified by the spectral clustering approach discussed in Section~\ref{UMFFR}. We further assume that an item can be part of the test set for a particular group if it is rated by $75\%$ of the users within the group, \adamya{and a maximum of 2 items can be included in the test set for each group}. It can be seen that $\{v_1, v_3, v_4, v_5\}$ meet the selection criteria for inclusion in the test set corresponding to group $\mathcal{U}^1$ whereas $v_4$ and $v_5$ satisfy the criterion for $\mathcal{U}^2$. The items $v_1$ and $v_4$ from group $\mathcal{U}^1$ and $v_5$ from group $\mathcal{U}^2$ are then randomly selected to the test set. We also ensure that each item must be rated by at least one group in the training set. A similar approach for train-test partition has been applied for recommendation tasks involving packages.  In package recommendation, for each package, only those users were selected in the test set who have rated at least $20\%$ of items of that package. For the package-to-group recommendation task, we first chose groups and packages with members $\le5$ and took those group-package combinations that have density $\ge50\%$. We limit the test set to include a maximum of 5 packages per group.

\subsection{Evaluation Metrics}
\label{sec:evalM}
In order to assess the recommendation accuracy of the models being compared, we have employed three notable evaluation metrics namely, Mean Absolute Error (MAE), Root Mean Square Error (RMSE), and Precision~\cite{herlocker2004evaluating}. Given a test dataset $Y^{test}_{n \times m}$ for $n$ users and $m$ items, and an index set of ground truth entries $\Omega$, for each ${i,j} \in \Omega$, the entry $y^{test}_{ij} \in \{1,2,...,R\}$, let $\hat{Y}^{test}_{n \times m}$ be the corresponding predicted rating matrix with entries $\hat{y}^{test}_{ij} \in \{1,2,...,R\}$. 

\vspace{0.1in}
\noindent MAE evaluates the arithmetic average of the absolute errors, i.e., the difference between the predicted value and the actual observed value whereas  RMSE calculates the square root of the average of squared errors. RMSE puts more emphasis on larger absolute errors by penalizing them heavily. Lower MAE and RMSE values correspond to higher accuracy of the recommender system. 

\begin{equation*}
    MAE = \frac{\sum \limits_{\{i,j\} \in \Omega}{|\hat{y}^{test}_{ij} - y^{test}_{ij}|}}{N}. \hspace{1cm} RMSE = \sqrt{\frac{\sum \limits_{\{i,j\} \in \Omega}{(\hat{y}^{test}_{ij} - y^{test}_{ij})^2}}{N}}.
\end{equation*}

\noindent Further, in most of the recommendation domains, the focus is on the prediction of top-$k$ most relevant items to a user. Precision \cite{zangerle2022evaluating} is one of the most popular metrics for the top-$k$ which refers to the fraction of recommended items that are relevant. For the $i$th user, precision can be computed as 
\[
Precision_i@k = \frac{d_i(k)}{k},
\]
where $d_i(k)$ is the number of relevant items in the top-$k$ recommendations for the $i$th user. We have evaluated the algorithms for the varying values of  $k \in \{1,2,3,4,5,10,20,30,40\}$.

The evaluation metrics discussed before are also applicable to recommendation tasks involving groups and packages. For example, in a group recommendation task, the prediction $\hat{y}^{test}_{ij}$ for $i$th user over $j$th item is set to the corresponding group prediction, and $y^{test}_{ij}$ corresponds to an entry in the test set. 

\subsection{Baselines}
\label{sec:baseline}


\adamya{
The major focus of our experimental analysis is to validate that the unified recommendation framework can be seamlessly integrated with most of the traditional matrix factorization-based collaborative filtering models, aiming to tackle diverse recommendation tasks simultaneously. To the best of our knowledge, there is no such matrix factorization-based unified framework that incorporates individual preferences over items along with group and package information to cater to various recommendation tasks. We have performed an extensive search to find baseline algorithms based on matrix factorization from each category that make use of individual preferences over items along with group and package information. We have reported the results related to baseline algorithms from individual recommendation categories in our experimental analysis to ensure a fair and comprehensive evaluation of our proposed approach. However, we have not found any matrix factorization-based algorithm for package-to-group that makes use of individual preferences over items along with group and package information. All the traditional approaches under this category are trained on datasets containing preferences of pre-defined groups over pre-defined packages. Therefore, we obtained benchmark results using the baseline algorithms, RMF and MMMF, wherein we generated preferences for a group (or package) by aggregating its corresponding users' (or items') preferences through mean aggregation. Subsequently, we compared these benchmark results against various proposed aggregation strategies outlined in Section~\ref{aggStrategies} for the package-to-group recommendation scenario.} 

For the personalized recommendation task, we have compared the results with the underlying MF models discussed in Section~\ref{MFCF}, namely, Regularized Matrix Factorization (RMF)~\cite{wu2007collaborative} and Maximum Margin Matrix Factorization (MMMF)~\cite{rennie2005fast}. \adamya{Additionally, we have also compared our models against three deep learning models stated as follows.}
\begin{itemize}
    \item DMF \cite{xue2017deep}: \adamya{Deep Matrix Factorization Model (DMF) makes use of deep neural architecture to implement matrix factorization-based collaborative filtering model. The authors proposed a novel cross-entropy loss function to handle the preference prediction over explicit ratings, unlike the other approaches where only implicit preferences are used to learn user and item representation.}
    \item VDMF and VNCF \cite{bobadilla2022vdmf}: \adamya{Deep Matrix Factorization (VDMF) and Variational Neural Collaborative Filtering (VNCF) models integrate DMF \cite{xue2017deep} and NCF \cite{he2017neural} with  Variational Autoencoder (VAE), respectively. The authors suggested a unified training process by combining the VAE's encoding and variational process with the regression layers of CF models to mitigate the conventional separate training methods.}
\end{itemize}

\noindent \adamya{To compare the efficiency of our framework in the context of group recommendation task,} we compared our algorithms with the following approaches tailored made for \adamya{group recommendation task}. 
\begin{itemize}
    \item AF, BF and WBF \cite{ortega2016recommending}: In this work, the authors \adamya{utilized} three different approaches to compute the group latent factors: (i) After Factorization (AF), (ii) Before Factorization (BF), and (iii) Weighted-BF (WBF). In AF, a least square-based matrix factorization model is used to compute the user and item latent factors. The group latent factor is obtained by merging the latent factors of all the users belonging to that group. In BF, a virtual user is created for each group, representing the item preferences of the group. The group factors are computed using the matrix factorization model over the virtual user. \adamya{In WBF, a weight is assigned to each item rated by the virtual user.}
    \item AOFRAM and AOFRAM\&W \cite{nam2021towards}:  The authors proposed an extended latent factor model which also takes into account the biases of items and users to obtain \adamya{fully} specified latent factor matrices. In \adamya{the} AOFRAM approach, if at least one group user rates an item, 
    the other group member's preferences \adamya{for that item} are filled using the learned latent factor matrices, else the item is used as a candidate for group recommendation. A virtual profile is created by averaging the observed and filled ratings for that item. \adamya{Similarly, virtual profiles are created for each group and item. Using the matrix factorization model, latent factors are learned for the virtual users, and preference prediction is made for 
    each group. Extending the AOFRAM model, AOFRAM\&W calculates a weighted average to derive the virtual user profile, where the weight for each group user is determined by the number of ratings given by that user.} 
\end{itemize}
For the package recommendation task, we have compared the performance of our proposed algorithms with the following approaches.  
\begin{itemize}
    \item $Min_{All}$ and $Mul_{All}$ \cite{wibowo2017matrix}: The authors obtain the latent factors by \adamya{applying} the matrix factorization model \adamya{across the entire} user-item rating matrix ($Min_{All}$ and $Mul_{All}$). \adamya{For each package, the package ratings are obtained either by considering the minimum of user ratings given to items within that package ($Min$) or by calculating the harmonic mean of user ratings for all items within the package ($Mul$).}     
\end{itemize}




\begin{figure*}[hbt!]
    \centering
        \begin{subfigure}[b]{0.49\textwidth}
            \centering
            \includegraphics[width=\textwidth, height= 6.5cm]{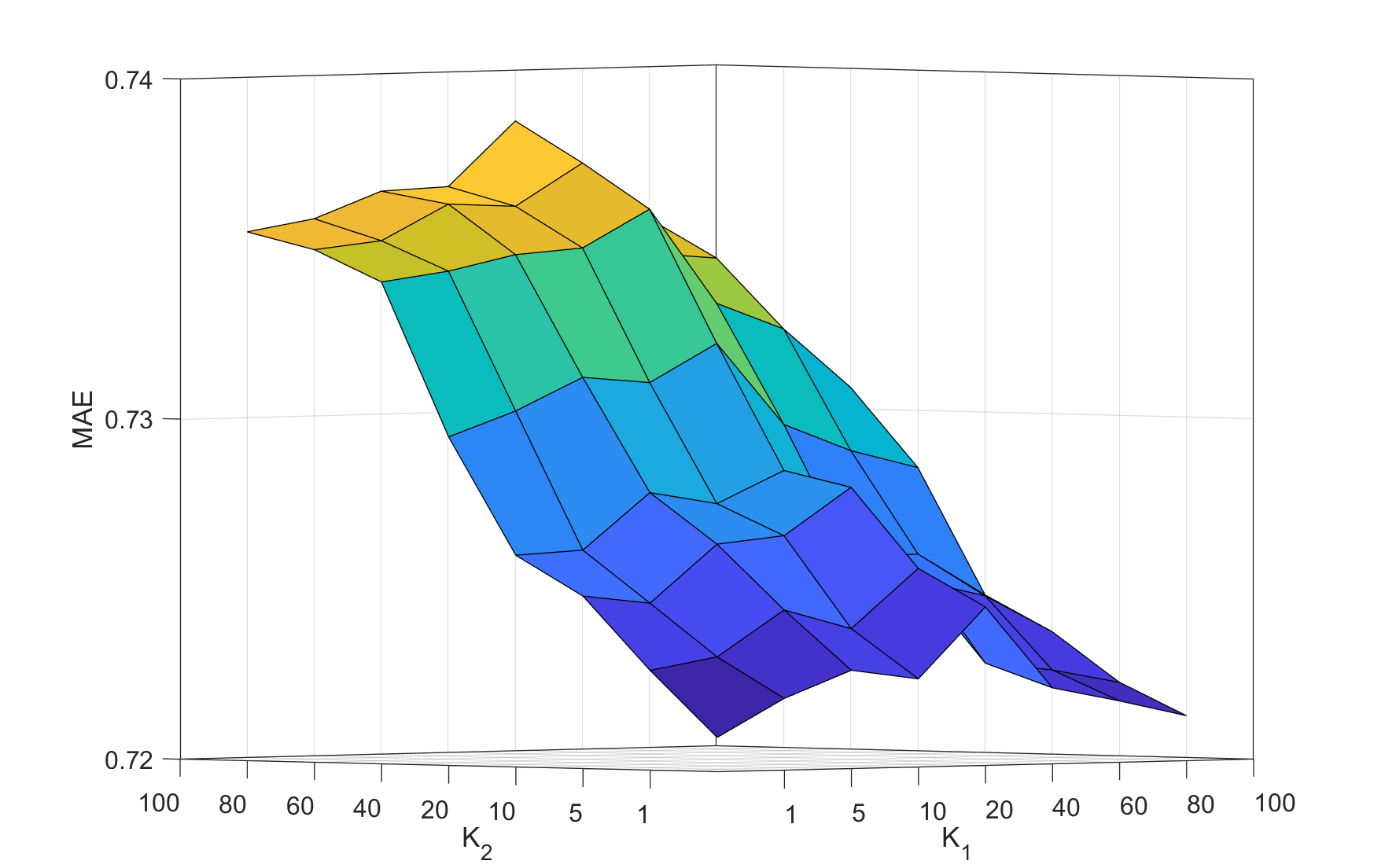}
            \caption{ Mean Absolute Error}
            \label{fig:UMMMFP100K}
        \end{subfigure}
        \hfill
        \begin{subfigure}[b]{0.49\textwidth}
            \centering
            \includegraphics[width=\textwidth, height= 6.5cm]{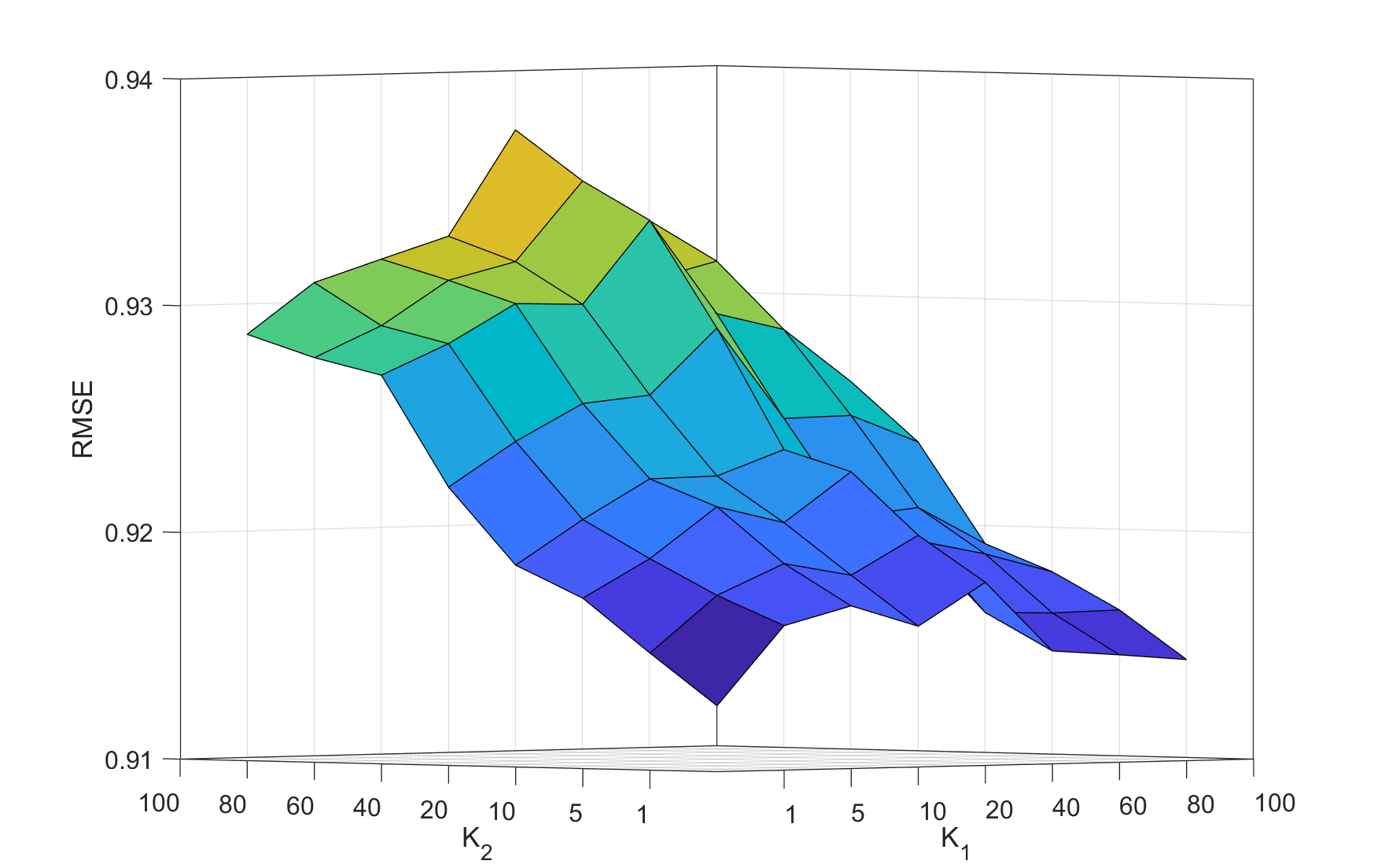}
            \caption{Root Mean Square Error}
            \label{fig:UMMMFGrp100K}
        \end{subfigure}
   \caption{Effect of $K_1$ and $K_2$ in URMF for personalized recommendation tasks over MovieLens 100K dataset.}
    \label{fig:UMFON100K}
\end{figure*}

\subsection{Parameter Configuration}
\label{sec:parameterTuning}
\adamya{To perform a comprehensive and fair evaluation, we have tuned the hyperparameters corresponding to each comparing algorithm in the following manner. The number of groups $K_1$ and packages $K_2$ is searched in $[1,5,10,20,40,60,80,100]$ for the personalized recommendation. In group and package recommendations scenarios, both $K_1$ and $K_2$ are searched in $[5,10,20,40,60,80,100]$. Finally, for the package-to-group recommendation, the number of members of a group or package is searched in $[2,4,6,8,10]$. The regularization parameters $\lambda$ and $\gamma$ in our proposed approach are searched in $\{i\:|\: i = 10^{j/16}, j \in \{0,1,2,...,23\}\}$. The regularization parameters for each baseline approach is tuned from $\{i\:|\: i = 10^{j/16}, j \in \{0,1,2,...,30\}\}$. The maximum number of iterations and the number of latent factors are set to be 200 and 100, respectively. For the deep neural network-based models DMF, VDMF, and VNCF, the batch size is searched in [32, 64, 128, 256], and the learning rate is searched in [0.0001, 0.001, 0.01, 0.1]. As reported in \cite{xue2017deep}, we have tuned the number of layers parameter in the range [2, 3] for DMF.  The optimal combination of hyperparameters for each comparing algorithm is determined by calculating the average of MAE, RMSE, and (1 $-$ Precision@$k$) for $k=[1,2,3,4,5]$ over the three runs. Subsequently, we selected the combination of hyperparameters corresponding to the minimum average score.}  
During the parameter tuning phase, we have observed slight variations in the scores of each evaluation metric across different combinations of $K_1$ and $K_2$.
Figure~\ref{fig:UMFON100K} shows the effect of varying $K_1$ and $K_2$ for personalized recommendation tasks using the unified RMF formulation (URMF) over the MovieLens 100K dataset. All the results reported in the subsequent section are obtained using fixed values of $K_1 = 20$ and $K_2 = 20$ for personalized, group, and package recommendation tasks, and for package-to-group recommendation, the value of  $K_1$ and $K_2$ is set to $\lfloor \frac{N}{8}\rfloor$ and $\lfloor \frac{M}{4}\rfloor$, respectively. We have observed similar comparative performance over other combinations of $K_1$ and $K_2$.

\subsection{Results and Discussion}
\label{sec:resultsandDiscussion}
This section presents the experiment results comparing the proposed approach with the other baseline algorithms over several categories of recommendation scenarios, including personalized, group, package, and package-to-group recommendations. 

\subsubsection{Personalized Recommendation}
\begin{figure*}[ht!]
    \centering
    \begin{subfigure}[b]{\linewidth}
        \centering
        \includegraphics[width=\linewidth, height= 4.9cm]{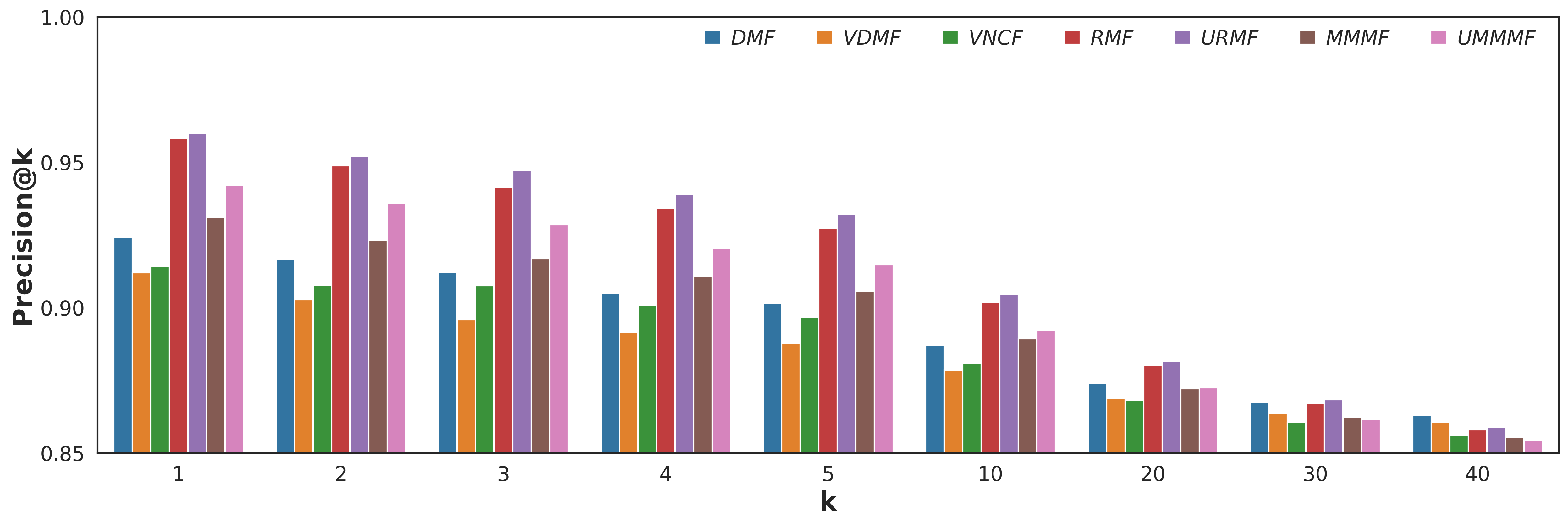}
        \caption{MovieLens 100K}
        \label{fig:Per100K}
    \end{subfigure}\hfill    
    \vspace{\baselineskip} 
    \begin{subfigure}[b]{\linewidth}
        \centering
        \includegraphics[width=\linewidth, height= 4.9cm]{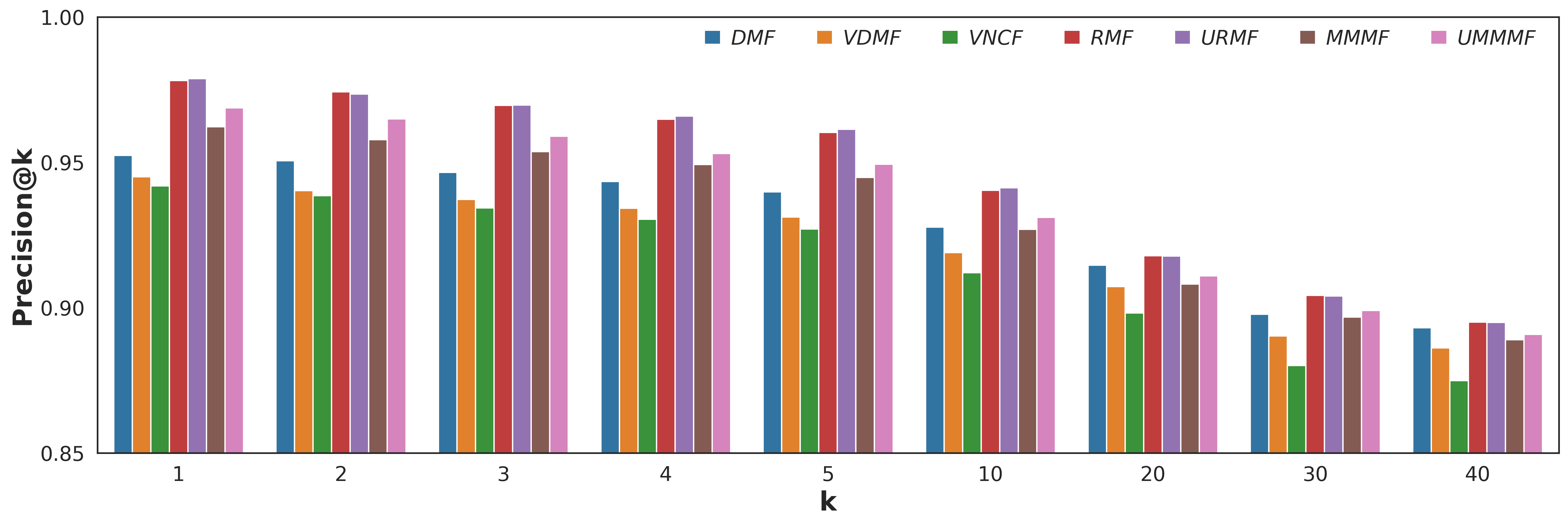}
        \caption{MovieLens 1M}
        \label{fig:Per1M}
    \end{subfigure}
    \caption{Precision@$k$ score of each comparing algorithm for Personalized Recommendation task.}
    \label{fig:PerPrecision}
\end{figure*}

\noindent The experimental results for personalized recommendation over MAE and RMSE metrics are shown in Table~\ref{Err:AccuPer}, where URMF and UMMMF refer to the unified framework corresponding to RMF and MMMF, respectively. 
\begin{table*}[ht!]
    \renewcommand{\arraystretch}{1.2}
    \centering
    \captionsetup{justification=centering}
    \caption{Experimental results of each comparing algorithm for Personalized Recommendation task.}
    \adjustbox{max width=0.7\linewidth}{
        \begin{tabular}{|l|l|c|c|c||c|c||c|c|} \hline
            \multicolumn{1}{|l|}{\textbf{Datasets}} & \textbf{Metrics} & \adamya{\textbf{DMF}} &\adamya{\textbf{VDMF}} & \adamya{\textbf{VNCF}} &\textbf{RMF} & \textbf{URMF} & \textbf{MMMF} & \textbf{UMMMF}\\ \hline
            \multirow{2}{*}{\textbf{MovieLens 100K}} & \textbf{MAE} & 0.7292 & 0.8485 & 0.7554 & 0.7672 & 0.7253 & 0.6868 & 0.6662 \\ \cline{2-9} 
             & \textbf{RMSE} & 0.9341 & 1.0805 & 0.9513 & 0.9792 & 0.9211 & 0.9899 & 0.9630 \\ \hline
            \multirow{2}{*}{\textbf{MovieLens 1M}} & \textbf{MAE} & 0.6835 & 0.7303 & 0.7353 & 0.7051 & 0.6779 & 0.6435 & 0.6245 \\ \cline{2-9} 
             & \textbf{RMSE} & 0.8749 & 0.9284 & 0.9269 & 0.8971 & 0.8593 & 0.9404 & 0.9164 \\ \hline
        \end{tabular}
        }
    \label{Err:AccuPer}
\end{table*}
\adamya{It can be seen that the proposed unified approach outperforms the corresponding underlying algorithm and other baseline algorithms across both datasets. 
Figure~\ref{fig:PerPrecision} illustrates Precision@$k$ scores for varying values of $k$. The unified framework demonstrates comparable or superior Precision@$k$ scores compared to the baseline algorithms, indicating higher predictive accuracy. Notably, the performance improvement of the unified model corresponding to the underlying algorithm highlights the importance of incorporating group and package information while modeling individual preferences. } 

\subsubsection{Group Recommendation}
\begin{figure}[ht!]
    \centering
    \begin{subfigure}[b]{\linewidth}
        \centering
        \includegraphics[width=\linewidth, height= 4.9cm]{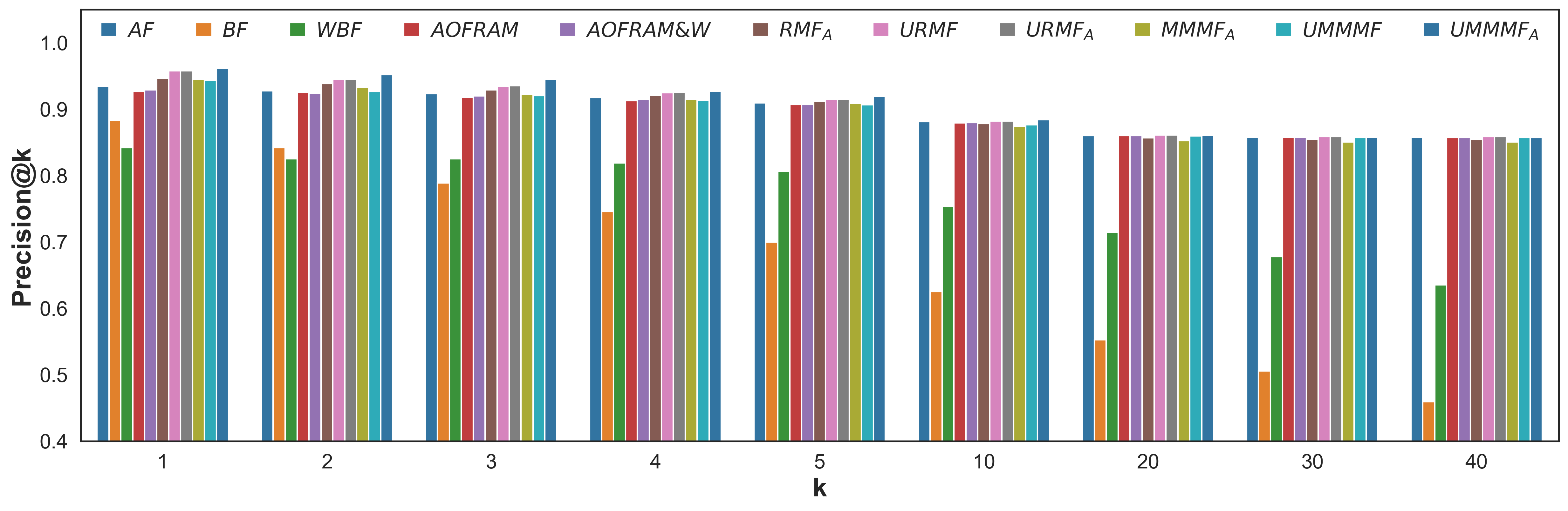}
        \caption{MovieLens 100K}
        \label{fig:Grp100K}
    \end{subfigure}
    
    \vspace{\baselineskip} 
    
    \begin{subfigure}[b]{\linewidth}
        \centering
        \includegraphics[width=\linewidth, height= 4.9cm]{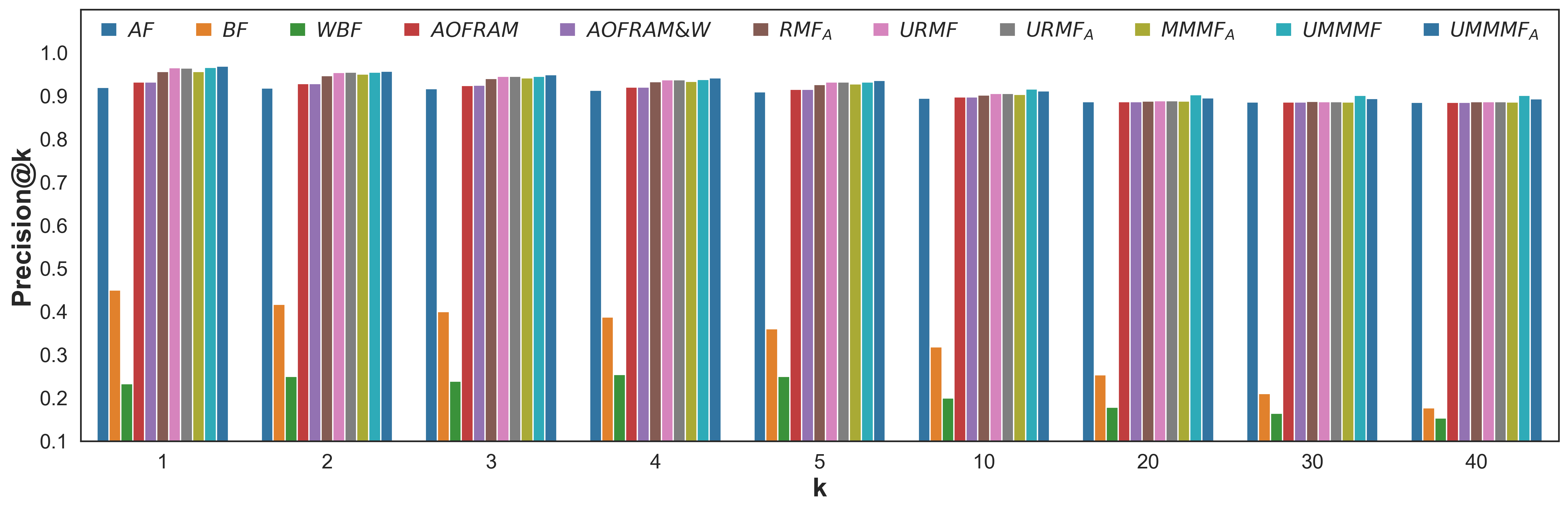}
        \caption{MovieLens 1M}
        \label{fig:Grp1M}
    \end{subfigure}
    
    \caption{Precision@$k$ score of each comparing algorithm for Group Recommendation task.}
    \label{fig:GrpPrecision}
\end{figure}

\noindent Table \ref{Err:AccuGrp} shows the comparison of our proposed approaches with baseline group recommendation models discussed in Section~\ref{sec:baseline} over MAE and RMSE measures. The subscript ``\textit{A}" with the model name is used to indicate that the group's rating for an item has been obtained by performing the mean-based aggregation of the predicted preferences of group members. 
\begin{table}[ht!]
    \renewcommand{\arraystretch}{1.5}
    \centering
    \captionsetup{font=scriptsize,justification=centering}
    \caption{Experimental results of each comparing algorithm for Group Recommendation task.}
    \adjustbox{max width=\linewidth}{
    \begin{tabular}{|p{2.1cm}|l|c|c|c|c|c||c|c|c||c|c|c|} \hline
        \multicolumn{1}{|l|}{\textbf{Datasets}} & \textbf{Metrics} & $\textbf{AF}$ & $\textbf{BF}$ & $\textbf{WBF}$ & $\textbf{AOFRAM}$ & $\textbf{AOFRAM\&W}$ & $\mathbf{RMF_A}$ & $\textbf{URMF}$ & $\mathbf{URMF_A}$ & $\mathbf{MMMF_A}$ & $\textbf{UMMMF}$ & $\mathbf{UMMMF_A}$ \\ \hline
        \multirow{2}{2.1cm}{\textbf{MovieLens 100K}} & \textbf{MAE} & 0.8139 & 0.8440 & 0.8927 & 1.0368 & 1.0509 & 0.8669 & 0.8013 & 0.8010 & 0.8107 & 0.7572 & 0.7828 \\ \cline{2-13} 
         & \textbf{RMSE} & 1.0082 & 1.0204 & 1.0991 & 1.3027 & 1.3133 & 1.0680 & 0.9977 & 0.9974 & 1.0869 & 1.0321 & 0.9972 \\ \hline
        \multirow{2}{2.1cm}{\textbf{MovieLens 1M}} & \textbf{MAE} & 1.0862 & 1.1237 & 1.5612 & 1.2164 & 1.2210 & 0.7873 & 0.7631 & 0.7627 & 0.7624 & 0.7099 & 0.7506 \\ \cline{2-13} 
         & \textbf{RMSE} & 1.4385 & 1.6165 & 1.9945 & 1.5025 & 1.4977 & 0.9705 & 0.9516 & 0.9512 & 0.9555 & 0.9891 & 0.9478 \\ \hline
    \end{tabular}
    }
    \label{Err:AccuGrp}
\end{table}
\adamya{It can be observed that the performance of the unified framework surpasses the performance of baseline methods, including its underlying approach by a notable margin. We have observed similar performance for the Precision@$k$ measure reported in Figure~\ref{fig:GrpPrecision}. This enhanced performance indicates the importance of capturing the intrinsic relationship between various recommendation tasks, particularly the crucial role of package information in learning effective group representations.}

\subsubsection{Package Recommendation}
\begin{figure}[ht!]
    \centering
    \begin{subfigure}[b]{\linewidth}
        \centering
        \includegraphics[width=\linewidth, height= 4.9cm]{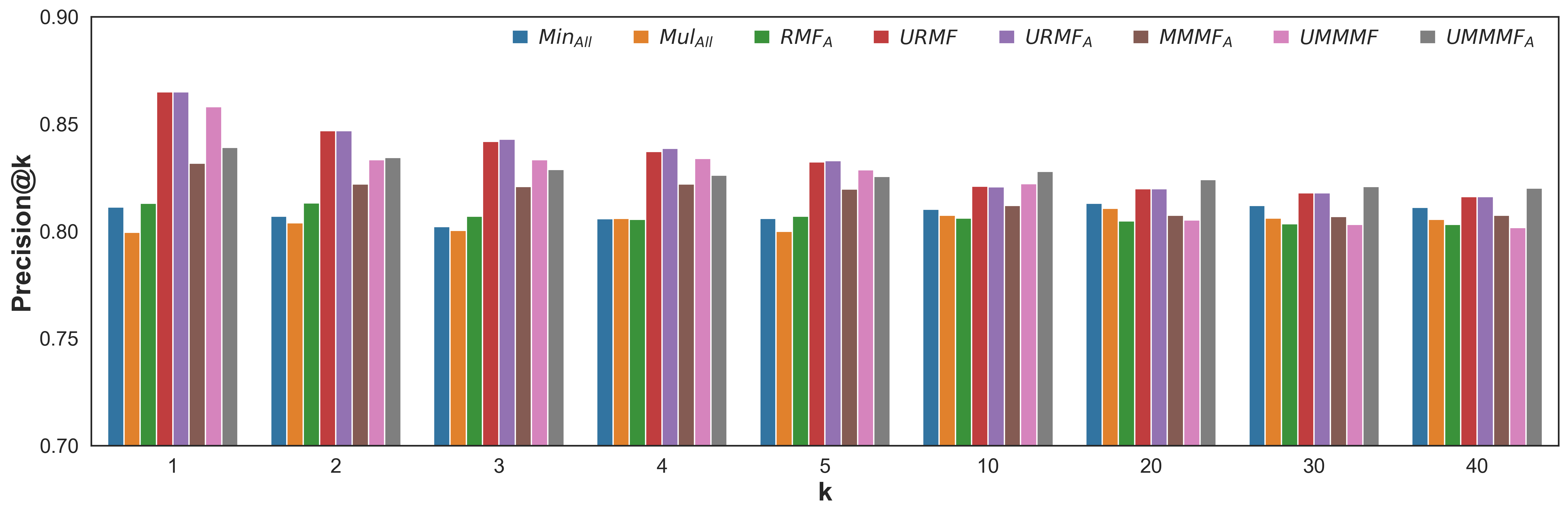}
        \caption{MovieLens 100K}
        \label{fig:Pkg100K}
    \end{subfigure}
    
    \vspace{\baselineskip} 
    
    \begin{subfigure}[b]{\linewidth}
        \centering
        \includegraphics[width=\linewidth, height= 4.9cm]{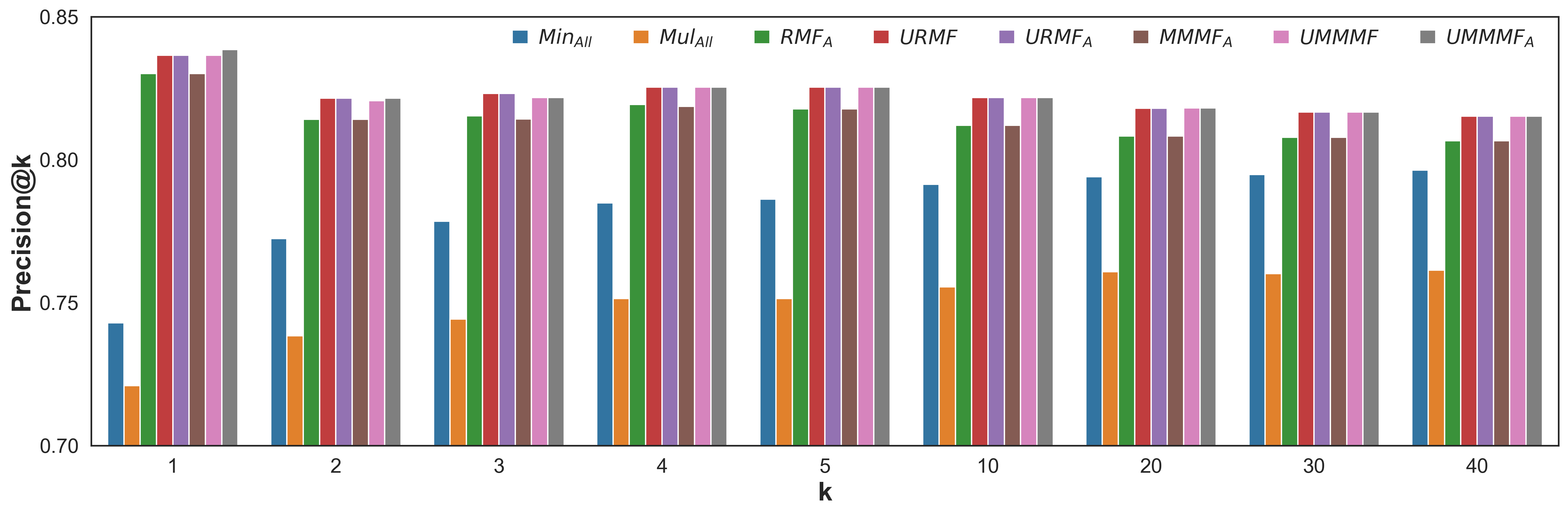}
        \caption{MovieLens 1M}
        \label{fig:Pkg1M}
    \end{subfigure}
    
    \caption{Precision@$k$ score of each comparing algorithm for Package Recommendation task.}
    \label{fig:PkgPrecision}
\end{figure}

\noindent 
The result related to the package recommendation task is reported in Table~\ref{Err:AccuPkg}, depicting the value for MAE and RMSE metrics. Subscript ``A" with the model name refers that the package rating for an individual user is derived through the mean-based aggregation of her predicted preferences for package items. It can be seen that the unified approach outperforms its counterparts across both evaluation metrics. We have observed similar performance in terms of Precision@$k$ measures reported in Figure~\ref{fig:PkgPrecision}. 
\begin{table*}
    \renewcommand{\arraystretch}{1.2}
    \centering
    \captionsetup{font=scriptsize,justification=centering}
    \caption{Experimental results of each comparing algorithm for  Package Recommendation task.}
    \adjustbox{max width=\linewidth}{
    \begin{tabular}{|l|l|c|c||c|c|c||c|c|c|}
        \hline
        \multicolumn{1}{|l|}{\textbf{Datasets}} & \textbf{Metrics} & $\mathbf{Min_{All}}$ & $\mathbf{Mul_{All}}$ & $\mathbf{RMF_A}$ & $\textbf{URMF}$ & $\mathbf{URMF_A}$ & $\mathbf{MMMF_A}$ & $\textbf{UMMMF}$ & $\mathbf{UMMMF_A}$ \\ \hline
        \multirow{2}{*}{\textbf{MovieLens 100K}} & \textbf{MAE} & 0.9877 & 0.8735 & 1.1071 & 0.8722 & 0.8722 & 0.8533 & 0.8362 & 0.7954 \\ \cline{2-10} 
        & \textbf{RMSE} & 1.1882 & 1.0577 & 1.3524 & 1.0920 & 1.0920 & 1.0880 & 1.1099 & 1.0192 \\ \hline
        \multirow{2}{*}{\textbf{MovieLens 1M}} & \textbf{MAE} & 1.5230 & 0.9575 & 1.9071 & 0.9412 & 0.9412 & 1.0669 & 0.9028 & 0.9181 \\ \cline{2-10} 
         & \textbf{RMSE} & 1.7751 & 1.1459 & 2.1593 & 1.1597 & 1.1598 & 1.3832 & 1.1631 & 1.1105 \\ \hline
    \end{tabular}
    }
    \label{Err:AccuPkg}
\end{table*}
\adamya{The results demonstrate the proposed framework as a promising solution for providing accurate and valuable package suggestions for users.}

\subsubsection{Package-to-Group Recommendation}
\begin{figure}[ht!]
    \centering
    \begin{subfigure}[b]{\linewidth}
        \centering
        \includegraphics[width=\linewidth, height= 4.9cm]{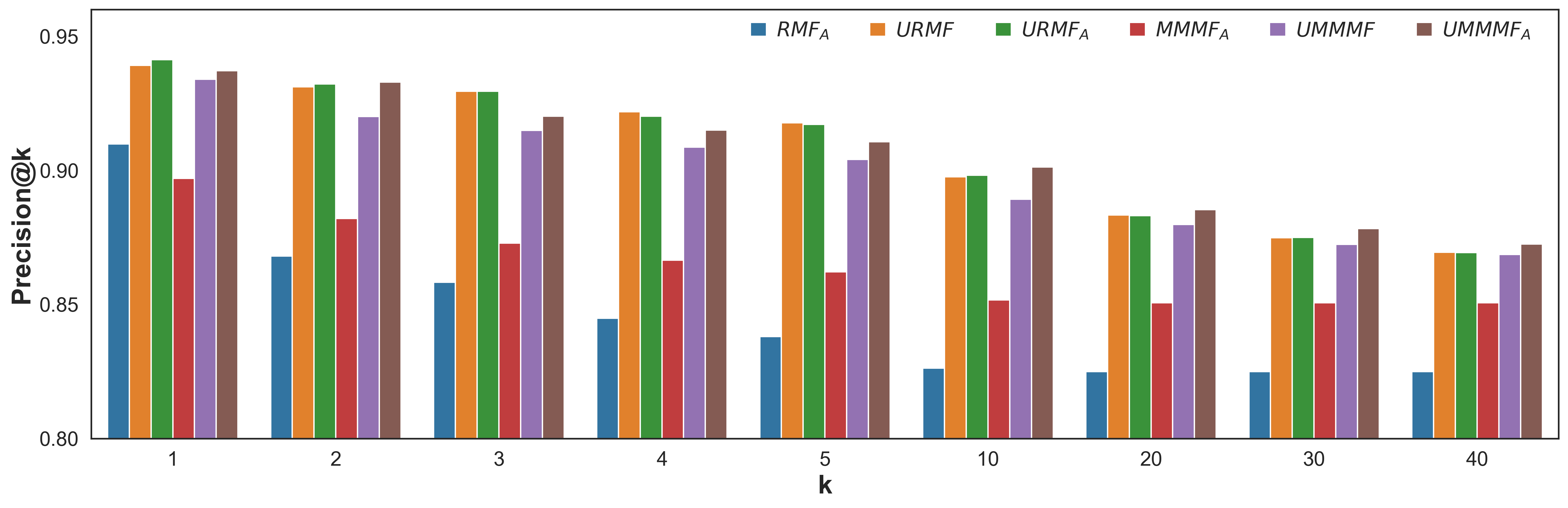}
        \caption{MovieLens 100K}
        \label{fig:P2G100K}
    \end{subfigure}
    
    \vspace{\baselineskip} 
    
    \begin{subfigure}[b]{\linewidth}
        \centering
        \includegraphics[width=\linewidth, height= 4.9cm]{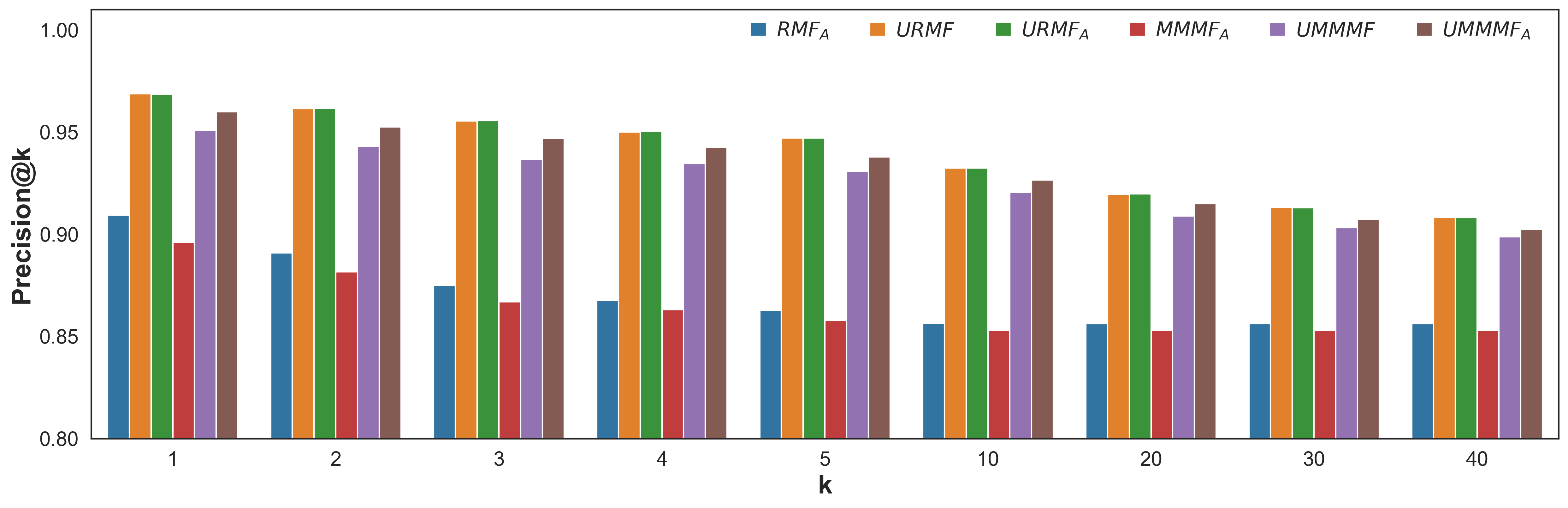}
        \caption{MovieLens 1M}
        \label{fig:P2G1M}
    \end{subfigure}
    
    \caption{Precision@$k$ score of each comparing algorithm for Package-to-Group Recommendation task.}
    \label{fig:P2GPrecision}
\end{figure}

\noindent  As mentioned previously, for this particular task, we did not find any existing algorithms that utilize user-item rating data. Therefore, we adopted the mean-based aggregation strategy to predict preference of users' and items' participating in the group and packages being assessed to obtain the group prediction over the package. Table \ref{Err:AccuP2G} shows the comparison of our proposed approaches with the mean-based aggregation strategy approaches for package-to-group recommendation scenarios. Subscript ``A" is used to refer results obtained by performing mean-based aggregation. 
\begin{table*}[hbt!]
    \renewcommand{\arraystretch}{1.1}
    \centering
    \captionsetup{font=scriptsize,justification=centering}
    \caption{Experimental results of each comparing algorithm for Package-to-Group Recommendation task.}
    \adjustbox{max width=\linewidth}{
    \begin{tabular}{|l|l|c|c|c||c|c|c|}
        \hline
        \multicolumn{1}{|l|}{\textbf{Datasets}} & \textbf{Metrics} & $\mathbf{RMF_A}$ & $\textbf{URMF}$ & $\mathbf{URMF_A}$ & $\mathbf{MMMF_A}$ & $\textbf{UMMMF}$ & $\mathbf{UMMMF_A}$ \\ \hline
        \multirow{2}{*}{\textbf{MovieLens 100K}} & \textbf{MAE} & 0.8541 & 0.8012 & 0.8009 & 0.7865 & 0.7323 & 0.7584 \\ \cline{2-8} 
         & \textbf{RMSE} & 1.0613 & 0.9954 & 0.9952 & 1.0275 & 1.0125 & 0.9728 \\ \hline
        \multirow{2}{*}{\textbf{MovieLens 1M}} & \textbf{MAE} & 0.8682 & 0.7667 & 0.7677 & 0.8625 & 0.7161 & 0.7428 \\ \cline{2-8} 
         & \textbf{RMSE} & 1.0722 & 0.9499 & 0.9506 & 1.0791 & 0.9905 & 0.9480 \\ \hline
    \end{tabular}
    }
    \label{Err:AccuP2G}
\end{table*}
The result demonstrates the efficacy of the proposed framework. The Precision@$k$ score for the same setting is reported in Figure~\ref{fig:P2GPrecision}. \adamya{Altogether, the results strengthen our claim that the unified framework achieves a superior level of accuracy compared with its counterparts in all four recommendation scenarios.} 


\adamya{The comprehensive analysis and results presented across multiple recommendation tasks underscore that the proposed unified framework can be simultaneously adopted for various recommendation scenarios such as personalized, group, package, and package-to-group recommendations. The efficiency of this unified approach across various evaluation metrics highlights its superior predictive accuracy. Further, the unified framework also signifies the importance of using group/package information as a guiding factor while learning the individual user/item representation in a personalized recommendation setting. The capability of the proposed approach to simultaneously learn the latent representation of users, items, groups, and packages emphasizes the strength and competitiveness of our proposed approach. The framework sets a foundation for future research endeavors to create more sophisticated and efficient recommendation algorithms.}

\section{Conclusion}
\label{sec:conclusion}

\adamya{
 The proposed work presents a novel unified recommendation framework that can be seamlessly integrated with most of the traditional matrix factorization-based collaborative filtering models. This unified framework jointly exploits representations of various entities like user, item, group, and package to facilitate their utilization across diverse recommendation scenarios involving these entities. We considered two baseline approaches, Regularized Matrix Factorization and Maximum Margin Matrix Factorization, and enhanced their formulation by incorporating components focusing on exploiting the group and package latent factors. The framework also provides flexibility to be easily integrated with matrix factorization-based personalized recommendation models employing many-objective optimization, including diversity, serendipity, and novelty for simultaneously learning the latent representation of users, items, groups, and packages~\cite{cui2021improved, symeonidis2019counteracting}. Experimental results over two real-world datasets demonstrated substantial performance improvements across various offline evaluation metrics for all four recommendation scenarios. The results also validate that the proposed approach captures intricate patterns and dependencies across different recommendation tasks, showcasing its application in various recommendation scenarios, ranging from e-commerce platforms to content streaming services and beyond. Furthermore, integrating group and package information while learning the user and item latent representations within personalized recommendation scenarios results in enhanced performance accuracy. This also validates that the group and package information can be used as an anchor point to guide the learning process, focusing on exploiting user and item representation.}
 
\adamya{
Despite the promising adaptability and potential of the proposed framework, there exist some limitations and areas for further improvement. Specifically, extracting group and package latent factors alongside user and item factors adds computational overhead, which could pose challenges for large-scale datasets.  In addition, the unavailability of datasets encompassing all possible preference combinations of users (groups) and items (packages) is also an issue.  The current findings are on datasets containing individual preferences over items, which are then processed to simulate groups and packages. However, this process of creating groups and packages might not fully capture the nuances and complexities of real-world group interactions, potentially impacting the generalizability and accuracy of the results. In the future, algorithmic improvements and model architectures able to handle large-scale data are areas to be explored, along with a focus on the fairness of the groups/packages formed. We also plan to explore various ways of incorporating side information, such as reviews, during the model building to learn explanatory factors aligned with the user, items, and group features for more accurate, transparent, and interpretable recommendations. Exploring applications of the proposed approach in a cross-domain recommendation setting with an emphasis on transferring the features learned for users (items) or groups (packages) is a direction worthwhile to explore.}
    
\section*{Acknowledgments}
 Vikas Kumar is supported by the Start-up Research Grant (SRG), Science and Engineering Research Board, India, under grant number SRG/2021/001931, UGC-BSR Start-up Grant, UGC, India, under grant number F.30-547/2021(BSR) and the Faculty Research Programme Grant, University of Delhi, under grant number IoE/2023-24/12/FRP. We would also like to thank the anonymous reviewers whose comments/suggestions helped to improve and clarify this manuscript to a large extent.

\bibliographystyle{unsrt}


\bibliography{references}

\end{document}